\documentclass[a4paper,10pt]{article}
\pdfoutput=1

\usepackage{jheppub} 
\usepackage[T1]{fontenc}
\usepackage{pstricks}
\usepackage{color}
\usepackage{graphicx}
\usepackage{multirow}

\title{Chances for SUSY-GUT in the LHC Epoch}

\author[a,b]{Zurab Berezhiani,}
\author[c,d]{Marco Chianese,}
\author[c,d]{Gennaro Miele}
\author[c,d]{and Stefano Morisi}

\affiliation[a]{Dipartimento di Fisica e Chimica, Universit\`a di L'Aquila, I-67100 Coppito, L'Aquila, Italy}
\affiliation[b]{INFN, Laboratori Nazionali del Gran Sasso, I-67010 Assergi, L'Aquila, Italy}
\affiliation[c]{Dipartimento di Fisica, Universit\`a di Napoli ``Federico II", Complesso Univ. Monte S. Angelo, Via Cinthia, I-80126 Napoli, Italy}
\affiliation[d]{INFN, Sezione di Napoli, Complesso Univ. Monte S. Angelo, Via Cinthia, I-80126 Napoli, Italy}

\emailAdd{zurab.berezhiani@aquila.infn.it}
\emailAdd{chianese@na.infn.it}
\emailAdd{miele@na.infn.it}
\emailAdd{stefano.morisi@gmail.com}

\abstract{
The magic couple of SUSY and GUT still appears the most elegant and predictive physics concept beyond the Standard Model. Since up to now LHC found no evidence for supersymmetric particles it becomes of particular relevance to determine an upper bound of the energy scale they have to show up. In particular, we have analyzed a generic SUSY-GUT model assuming one step unification like in SU(5), and adopting {\it naturalness} principles, we have obtained general bounds on the mass spectrum of SUSY particles. We claim that if a SUSY gauge coupling unification takes place, the lightest gluino or Higgsino cannot have a mass larger than  $\sim 20$ TeV. Such a limit is of interest for planning new accelerator machines.}

\begin{document}
\maketitle
\flushbottom

\section{Introduction}

The recent detection in July 2012, of a new particle compatible with the Higgs boson  by ATLAS \cite{Aad:2012tfa} and CMS \cite{Chatrchyan:2012ufa} collaborations at LHC certainly represents a milestone that has once again confirmed the predictive strength of the Standard Model (SM) of strong and electroweak interactions.  However, from the theoretical point of view it is hardly to believe that the SM is the last step toward the unification in a simple principle of all the fundamental interactions. Several phenomena or open problems suggest the presence of physics beyond the SM. One can just remind few of them, like the hint of unification of all gauge couplings for extreme large energy, the particular structure of fermion masses, the baryonic/leptonic number violation processes necessary at scale larger than the electroweak one in order to yield baryogenesis, the problem related to the separation of very different energy scales in a field theory with scalars ({\it hierarchy problem}),  etc.

The Grand Unified Theory (GUT) paradigm is able to address part of these problems. Suggested by the simple extension of the SM scheme to larger gauge groups, it has represented a huge scientific effort, which, starting from seventies, has had to face with the experimental counterpart  of proton decay search, and with the necessity of an even too rich scalar content of the theory. The main feature that makes plausible a GUT scheme at very high energy, namely a compact  gauge group describing the fundamental interactions, comes from the observation that running gauge couplings tend to get closer  and closer with the increase of energy. 

A GUT is generally characterized  by a typical energy scale, $M_{\rm GUT}$. It is naively defined as the scale where the SM gauge couplings cross at one point according to their running.  The same definition can  be rephrased as follows: the GUT scale is the scale at which a compact gauge symmetry $G$ spontaneously breaks down to SM  (namely $SU(3) \times SU(2) \times U(1)$). Below such a scale the gauge couplings, being equal to each other at $\mu = M_{\rm GUT}$, run down and match  their experimental values at the benchmark point $\mu = M_Z$. However, this is strictly true for $SU(5)$ model only, for which both the above definitions are equivalent. Moreover, in $SU(5)$ the {\it great desert} is occurring since there are no intermediate thresholds due to new massive states  starting from $M_t$ up to GUT scale.

Then a question is in turn and concerns the size of $M_{\rm GUT}$. While a natural upper bound is represented by string or reduced Planck scale $M_P\simeq 2 \cdot 10^{18}$~GeV, the lower limit is settled by the predicted degree of stability of ordinary matter (proton decay). The experimental limits on the proton decay mediated by the heavy gauge bosons pose stringent bounds on their masses, hence on $M_{\rm GUT}$.

Before LEP epoch, non-supersymmetric GUTs could not been discriminated due to the poor precision characterizing the measurements of SM parameters, even though the low energy scale of unification, order $\lesssim 10^{15}$~GeV, was already in strong tension with the proton lifetime limit. Such situation  then drastically changed with the precise measurements at LEP, which essentially excluded non-supersymmetric GUT. In fact, after the precise measurements of the gauge couplings $\alpha_1$, $\alpha_2$, $\alpha_3$ at $M_{\rm Z}$, and of other SM quantities at LEP, it was clear that a real one step unification of running gauge couplings was not possible at high energy, at least without assuming new physics. However, in the framework of Supersymmetry (SUSY) one obtains a milder running of couplings that provides a satisfactory unification at a common high energy scale ${M_{\rm GUT}}$  \cite{Dimopoulos:1981yj, Ibanez:1981yh, Einhorn:1981sx, Marciano:1981un, Amaldi:1991cn}.  Moreover, SUSY also provides a road map for solving the gauge hierarchy and Doublet-Triplet splitting problems. In particular, minimal SU(5) allows a technical solution for the hierarchy and Doublet-Triplet splitting {\it via} a fine tuning, which is stable against radiative corrections due to SUSY \cite{Dimopoulos:1981zb, Sakai:1981gr}.  However, there exist more {\it natural} solutions without fine tuning,  as ``Missing Doublet Mechanism" (MDM) in extended SU(5) \cite{Georgi:1981vf, Masiero:1982fe, Grinstein:1982um, Hisano:1994fn, Berezhiani:1996nu}, ``Missing v.e.v.  Mechanism" (MVM) in SO(10) \cite{Dimopoulos:1981xm, Srednicki:1982aj, Babu:1994dq, Berezhiani:1996bv},  or ``pseudo-Goldstone boson instead of Fine Tuning" (GIFT) Mechanism in SU(6) \cite{Berezhiani:1989bd, Barbieri:1993wz, Berezhiani:1995sb, Barbieri:1994kw,Berezhiani:1995dt,Dvali:1996sr}.\footnote{The rudimentary idea of Higgs emerging as pseudo-Goldstone boson was first introduced at the level of SUSY-$SU(5)$ model in Refs. \cite{Inoue:1985cw,Anselm:1986um, Anselm:1988ss}, assuming that Higgs superpotential has an {\it ad hoc} global $SU(6)$ symmetry at the price of introducing an extra singlet supermultiplet. These models are not fully realistic without local $SU(6)$ completion. At the level of local $SU(5)$ symmetry the assumption of  Higgs superpotential  having $SU(6)$ global symmetry is equivalent to many fine tunings, but at the same time such extended symmetry is  explicitly violated in the Yukawa sector. Let us remark that the supersymmetric pseudo-Goldstone mechanism was also discussed in the context of $SO(10)$ \cite{Barbieri:1992yy}, $SO(n)$ and $E_6$  \cite{Berezhiani:1995sb}. More recently such mechanism was also applied for the solution of the little hierarchy problem in the framework of minimal low scale extensions of the MSSM \cite{Berezhiani:2005pb,Roy:2005hg,Csaki:2005fc,Berezhiani:2005ek,Falkowski:2006qq}.}
  
 The golden goal of supersymmetry was to explain naturally the origin of the electroweak (EW) scale in terms of the soft supersymmetry breaking (SSB) parameters, without artificial fine tunings among them. In this design of a {\it natural} SUSY, the EW scale has to be originated by the SUSY breaking scale itself.  This means that already above few hundred GeV the SM has to be replaced by the supersymmetric extensions of the Standard Model (MSSM in the minimal version),  which would imply a rich new phenomenology that LHC should have already found. With the conclusion of the 8 TeV LHC run I, MSSM have been greatly constrained by a variety of direct searches \cite{Agashe:2014kda,ATLAS-webpage,CMS-webpage}.  The properties of the discovered Higgs boson are fully compatible with the Standard Model Higgs,  and a second light Higgs predicted in natural SUSY was not found. In the meanwhile the experimental lower bounds on the masses of gluinos and squarks were increased up to TeV scale.  

More in general, the missing evidence up to now for a SUSY phenomenology at LHC has caused a pessimistic attitude of the scientific community toward SUSY  and SUSY-GUT paradigms, thus forcing theoretical physicists to look for different approaches.  As a radical possibility, there emerged a concept of split supersymmetry \cite{Giudice:2004tc, ArkaniHamed:2004yi},  which pushes some SUSY partners (squarks and sleptons) to arbitrarily large scale while leaving the others (gauginos and Higgsinos) at the low scale.  
 
In view of this, one can ask if the experimental situation after LHC is really so dramatic to justify almost a total give up of the SUSY scheme, or SUSY still remains the most promising and well defined concept that particle physics above SM should tend to. In any case one cannot exclude the possibility that SUSY lives just around the corner, and gluinos or squarks will be indeed discovered at the TeV scale in the second run of the LHC.   However, it is important to remark that even before LHC, already in the LEP epoch  the experimental data supplied mounting evidences disfavoring, at some level, SUSY at few hundred GeV  \cite{Barbieri:2000gf}.  It is worth reminding the problem of electric dipole moments of electron and neutron, which within $\sim$100 GeV range for SUSY were predicted too large, while increasing the SUSY scale to few TeV they are  naturally suppressed  to the level of the present experimental limits.  The situation is similar for the flavor violating processes like  $\mu \to e\gamma$, $b\to s \gamma$ etc.  For arbitrary soft parameters, the flavor violation limits  would require the increasing of the SUSY scale up to 100 TeV, but such a limit can be lowered to TeV scale by a particular use of symmetry arguments, namely by means of an approximate alignment of the soft SUSY parameters with the Yukawa couplings. In this concern, one can mention particular theoretical scenarios based on a gauged family symmetry $SU(3)_H$  \cite{Berezhiani:1996ii, Berezhiani:1996kk, Anselm:1996jm, Berezhiani:2001mh}, which can be implemented for  SUSY-GUT as well  \cite{Berezhiani:1996ii, Berezhiani:1996kk, Berezhiani:2005tp}. In this case, the flavor violation limits would allow SUSY scales as small as 1 TeV.  Later on such a possibility was described as a paradigm thereafter simply denoted as Minimal Flavour Violation \cite{D'Ambrosio:2002ex}. 

Finally, as will be clarified in the following,  the Higgs physics, modulo specific conspiracies, seems to indicate that the SUSY scale should be above few TeV as well. Therefore, one can conclude that SUSY scale larger than few TeV is quite reasonable, and if SUSY indeed lives near this lower bound, then it should show up in the LHC run II. 

In this paper, we try to reanalyze the issue showing the room still remaining for SUSY-GUT inspired models under some natural assumptions, which will be clarified in the following, and the consequences of such schemes on LHC run II and on future colliders like FCC-ee and FCC-hh \cite{Gomez-Ceballos:2013zzn, fcc_hh}. Under such assumptions, we study the implications on SUSY mass spectrum due to the gauge couplings and Yukawa couplings unification. We adopt the paradigm of one step gauge unification, which we simply denote as {\it SU(5) bottleneck}, and that still remains a quite general condition. GUTs may be based on a group larger than $SU(5)$  like $SO(10)$, $SU(6)$ or $E_6$, which provide promising and predictive models describing physics below the string scale or reduced Planck scale $M_P$.  Such latter symmetry groups do not provide automatically  one step unification of the gauge couplings, and generically they could be broken down to  the SM passing through different intermediate stages of the gauge symmetry breaking. In this case the gauge couplings unification phenomenon would not be a clean prediction of the scheme, but rather an {\it ad hoc} reconstructed phenomenon. In this respect, the condition of {\it SU(5) bottleneck}  simply requires that the gauge symmetry is broken in such a way that below the scale $M_{\rm GUT}$  it reduces to $SU(3) \times SU(2) \times U(1)$ with three gauge couplings $\alpha_1$, $\alpha_2$, $\alpha_3$ that run down in energy starting from {\it the same} value $\alpha_{\rm GUT}$  (denoting the GUT gauge coupling at $M_{\rm GUT}$).  In other words we require that independently of whatever is the ``mother" unified gauge group, and the intermediate breaking scales it may require, at the last symmetry breaking stage,  nearby $M_{\rm GUT}$, it behaves as $SU(5)$.  Below such a scale the theory should reduce to the Minimal Supersymmetric Standard Model (MSSM).

In the present analysis we require that all coupling constants, including Yukawa couplings, must be order one in the ``mother" GUT scenario above $M_{\rm GUT}$, assuming that there are no {\it ad hoc} small parameters and no {\it ad hoc} fine tuning among parameters. Moreover, the mass parameters involved in the SUSY breaking are supposed to be of the same order of magnitude (modulo possible differences between $F$- and $D$-terms).  Small Yukawa couplings for light families in the MSSM superpotential below the GUT scale  do not contradict the previous requirements as we will discuss in the next section. 

In summary, to perform our study we require:  
\begin{itemize}
\item Unification of all gauge couplings at a single energy scale ($M_{\rm GUT}$) without intermediate symmetry scales ({\it SU(5) bottleneck}). We take into account only SUSY particle thresholds, simply denoted as SUSY thresholds, at which SUSY particles show up, and GUT thresholds related to GUT multiplet fragments that can be lighter than $M_{\rm GUT}$. Since we do not allow the presence of {\it ad hoc} small parameters these fragments cannot be much lighter than one order of magnitude of $M_{\rm GUT}$.  
\item Consistency of third family fermion masses: correct mass of the top quark and  Yukawa $b$-$\tau$ unification. 
\item Consistency with the experimental limit on proton decay.
\item The absence of special fine tunings among the parameters (couplings ${\cal O}(1)$ at $M_{\rm GUT}$), which can be seen as a general {\it naturalness} requirement.
\end{itemize}
Moreover, we do not consider the possible limits on SUSY-GUTs coming from the assumption that one of the SUSY particle must necessarily be a DM candidate. This is due to the possible presence of R-violating terms that would make unstable the lightest SUSY particle at cosmological time-scale. Hence, in order to get model independent predictions we prefer do not use such cosmological constraints. 

Under this ansatz, we analyze the limits on the mass spectrum of SUSY particles once the previous requirements are fulfilled, and indeed we obtain an upper limit of about  few TeV in the more natural case when all supersymmetric particles have masses of the same order of magnitude. The situation changes if one takes into account a possible spread among the superpartner masses of different types (gauginos, sfermions and Higgsino). In this case, we look for the minimum of the mass spectrum, identifying such value as the energy scale upon which SUSY phenomenology has to be detected. By spanning on all compatible SUSY-GUT models (see previous requirements), in presence of both SUSY and GUT thresholds, we find that the above minimum can be as large as $\sim 20$ TeV. This represents the main result of our analysis since it provides an almost model independent upper bound for the appearance of  SUSY phenomenology.

In order to get such a result, once that all masses of the model are fixed, by using a \textit{Mathematica} code that solves the set of Renormalization Group Equations (RGEs) at two loops, we determine the unification GUT scale $M_{\rm GUT}$ and the possible compatibility of such a choice with the experimental measurements of gauge couplings at EW scale. Note that the running of Yukawa couplings is also taken into account as well as the constraints coming from the top mass and possible b-$\tau$ unification requirement. Finally, from this approach we get the allowed regions for the SUSY particle masses determined by the simultaneous effect of SUSY and GUT thresholds. From these results one obtains indications about the discovery potential of SUSY at LHC  hence clarifying the role of future colliders. Note that the effect of possible intermediate thresholds could be also studied, but it would result in an extremely model dependent scenario. For this reason in order to be more predictive we prefer do not consider such possibility. 

The paper is organized as follows. In section II we give the generic details of SUSY-GUT model considered, whereas in section III we provide the Renormalization Group Equations for couplings, discussing the initial conditions and SUSY, GUT thresholds as well. Moreover this section contains a description of the numerical method adopted. In section IV we report our results whereas section V contains our conclusions.

\section{Overview on SUSY-GUT and naturalness principles}

Let us consider a generic SUSY-GUT based on a gauge symmetry $G$. Within the framework of $N=1$ supersymmetry, such a theory should contain vector (gauge) superfields $V$ in adjoint representation of $G$, and some set of chiral superfields $\Phi$ in different representations of $G$.  A generic renormalizable Lagrangian can be written as\footnote{For  explicit notations, see e.g. Ref. \cite{Martin:1997ns}.} 
\begin{equation}
\mathcal{L}_{\rm SUSY} = \int d^2 \theta d^2 \bar{\theta}\, \Phi^{\dagger} e^{V}\Phi + 
\left[\int{d^2\theta \, \mathcal{W} \mathcal{W}} + \int{d^2 \theta \, W(\Phi)} + {\rm h.c.}\right] \,,
\label{eq:susy-lagrangian}
\end{equation}
where the first and second term yield the canonically normalized  kinetic terms of the chiral superfields and the gauge superfields, and their gauge interactions. These terms do not involve any coupling constant apart of the gauge coupling, which is an $\cal{O}$$(1)$ parameter.   Third term describes mass and interaction terms between the fermionic and scalar components of chiral superfields $\Phi$, via the superpotential $W(\Phi)$ that  is a general $G$ invariant holomorphic combination $\Phi$ containing the trilinear terms with $\cal{O}$$(1)$ coupling constants and the bilinear (mass) terms. The soft supersymmetry breaking (SSB) terms can be presented in a form similar to Eq.~\eqref{eq:susy-lagrangian} 
\begin{equation}
\mathcal{L}_{\rm SSB} = \int d^2 \theta d^2 \bar\theta\, \rho \, \Phi^{\dagger} e^{V}\Phi + 
\left[ \int{d^2\theta \, \eta \, \mathcal{W} \mathcal{W}} + \int{d^2 \theta \, \eta \, W'(\Phi)}+ {\rm h.c.}\right] \,,
\label{eq:ssb-lagrangian}
\end{equation}
making use of auxiliary superfields with non-zero $F$- and $D$-terms, respectively, $\eta = M_F \, \theta^2$ and $\rho = M_D^2 \, \theta^2\bar{\theta}^2$, where the dimensional parameters $M_F$ and $M_D$ can be in principle different.  In particular, $M_F$ determines the size of gaugino mass terms, and via the third term in Eq. \eqref{eq:ssb-lagrangian}, contributes to the SSB terms for the chiral superfields. The $D$-term can be of the order of $M_F^2$ once it is simply given by a direct product $\eta \bar\eta = M_F^2 \theta^2 \bar{\theta}^2$,  while there can be also a direct $D$-term and in this case $M_D^2 \gg M_F^2$. In general the function $W'(\Phi)$  should have the same structure of the superpotential $W(\Phi)$, but its couplings are not obliged to be the same of $W(\Phi)$.

The chiral superfields can be divided into Higgs and fermion superfields distinguished by matter parity $Z_2$, under which the fermion superfields change the sign while the Higgs superfields are invariant. In this way, the superpotential $W$ has two terms related to the Higgs and Yukawa sectors, namely
\begin{equation}
W=W_{\rm Higgs}+W_{\rm Yukawa} \,.
\label{eq:superpotential}
\end{equation}
The Higgs superpotential  $W_{\rm Higgs}$ is responsible for the v.e.v.'s breaking both the gauge symmetry $G$ down to MSSM and then to SM. 

\subsection{MSSM limit of a SUSY-GUT: mass scales and Higgs sector}

The MSSM represents the low energy limit of a generic SUSY-GUT. Here, we are going to discuss the different energy scales entering in the SUSY breaking pattern, paying particular attention to the Higgs sector. In MSSM the superpotential of Eq. \eqref{eq:superpotential} is explicitly given by
\begin{equation}
\label{eq:W-MSSM}
W_{\rm MSSM} =
Y^u _{ij} Q_i u^c_j H_u + Y^d _{ij} Q_i d^c_j H_d + Y^e _{ij} e^c_i L_j H_d  +  \mu H_u H_d \,,
\end{equation}
which contains chiral superfields corresponding to three families of quarks and leptons ($i,j=1,2,3$ are family indices), and two Higgses $H_u$ and $H_d$. The SSB $F$-terms repeat the structure of $W_{\rm MSSM}$, and contain also the soft Majorana masses of gauginos (bino, neutralinos and gluinos for $a=1,\,2,\,3$)
\begin{equation}
\label{eq:F-MSSM}
\mathcal{L}_{\rm F}= A^u _{ij} \tilde{Q}_i \tilde{u}^c_j H_u + A^d _{ij} \tilde{Q}_i \tilde{d}^c_i H_d + A^e _{ij}  \tilde{e}^c_i \tilde{L}_j H_d + \mu B_\mu H_u H_d + \tilde{m}_{\rm G}^a \lambda_a \lambda_a\,.
 \end{equation}
Note that the dimensional parameters $A^{u,d,e}_{ij}$, $B_\mu$ and $\tilde{m}_{\rm G}^a$ are expected to be parametrically of the order of $M_F$, though in the model dependent context the gaugino masses can be in principle different from other $F$-terms. On the other hand, the soft masses of all scalars including the Higgses are given by $D$-terms 
\begin{equation}
\label{eq:D-MSSM}
\mathcal{L}_{\rm D}= \tilde{m}^2_{Qij} \tilde{Q}^\dagger_i \tilde{Q}_j + 
\tilde{m}^2_{uij} \tilde{u}^{c \dagger}_i \tilde{u}^c_j + \tilde{m}^2_{dij} \tilde{d}^{c \dagger}_i \tilde{d}^c_j + \tilde{m}^2_{Lij} \tilde{L}^\dagger_i \tilde{L}_j + \tilde{m}^2_{eij} \tilde{e}^{c \dagger}_i \tilde{e}^c_j + \tilde{M}^2_u H^*_u H_u + \tilde{M}^2_d H^*_d H_d \,.
 \end{equation}
All these dimensional parameters are of the same order of magnitude given by the energy scale $M_D$.  These terms, in principle, can be parametrically larger than $F$-terms, though they can naturally be  of the same order of magnitude.  Therefore, all soft parameters can be divided in three classes: the soft Majorana gaugino masses $\cal{O}$$(M_F)$, the soft masses of the scalars as squarks and sleptons $\cal{O}$$(M_D)$, and the so-called $\mu$-term that determines the Higgsino masses and contributes to masses of scalar doublets $H_u$ and $H_d$. GUT implies that all gauginos must have the same mass at the GUT scale ($\tilde{m}_{\rm G}^a= \tilde{m}_{\rm G}$ $\forall a$), and the similar mass unification can be assumed for masses of the squarks and sleptons entering in the same GUT multiplet. Hence the differences between the true masses of squarks and sleptons, or between the masses of gluinos and neutralinos, are simply due to the running. In this way, SUSY-thresholds, namely the possible splittings among the masses of SUSY particles, can be defined by three mass scales only, which are gluino mass $\tilde{m}_g$ $\cal{O}$$(M_F)$, squark mass $\tilde{m}_{sq}$ $\cal{O}$$(M_D)$, and the Higgsino mass $\tilde{m}_H$ $\cal{O}$$(\mu)$.

A comment concerning the Higgs scalars $H_u$ and $H_d$ is in turn. Their mass matrix  
\begin{equation}
\label{eq:Hmatrix}
{\cal M}^2 = \left( \begin{array}{cc}
\tilde{M}^2_u + \mu^2 & \mu B_\mu \\
\mu B_\mu & \tilde{M}^2_d + \mu^2
\end{array} \right)
\end{equation}
involves the mass parameters of three different origins: mass terms $\tilde{M}^2_u$ and $\tilde{M}^2_d$ (SSB $D$-terms),  $B_\mu$ (SSB $F$-term), and  supersymmetric $\mu$-term, which can be of different orders of magnitude. If the SSB parameters assume large values, namely larger than few TeV or even larger, one cannot pretend to recover the so-called {\it natural solution}, which would link the EW scale to the SSB scale. Hence a certain fine tuning condition has to be imposed. In particular, one eigenstate $h$ should have a small negative squared mass
\begin{equation}
\label{eq:Higgs_finetunig}
- m^2  = \frac 12 \left( 2 \mu^2 + \tilde{M}^2_u + \tilde{M}^2_d - \sqrt{4 \mu^2 B_\mu^2 +(\tilde{M}^2_u - \tilde{M}^2_d)^2}\right) \sim - (100~{\rm GeV})^2 \,,
\end{equation}
and that will be identified with the SM Higgs. It should get a v.e.v.  ($v = 250$~GeV) breaking the EW symmetry and leaving behind the Higgs Boson with mass $M_h^2  = 2 m^2 \approx (125~{\rm GeV})^2$. The other mass eigenstate $\tilde h$ is heavy, with $M^2_{\tilde h} = 2 \mu^2 + \tilde{M}^2_u + \tilde{M}^2_d + m^2$, and it should decouple at the SSB scale. In this way, the mixing angle between the Higgses $H_u$ and $H_d$ is given by $\tan 2\beta = 2\mu B_\mu/(\tilde{M}^2_u - \tilde{M}^2_d)$, and hence the v.e.v. of $h$ is placed between them $v_u = v\sin\beta$ and $v_d = v\cos\beta$. 

In principle, $D$-terms $\tilde{M}_{u,d}$ could be larger than $B_\mu$ and $\mu$. However in this case the fine tuning of Eq. \eqref{eq:Higgs_finetunig} would not be possible. On the other hand, a $F$-term $B_\mu$ much larger than $D$-terms is not very natural. Moreover, for $\mu, B_\mu \gg \tilde{M}_{u,d}$ one would get $\tan \beta\approx 1$, which would make problematic  to accommodate the 125~GeV Higgs. Therefore, to impose a realistic fine tuning only one possibility is left, namely to take soft parameters $\tilde{M}_{u,d}$, $B_\mu$ and supersymmetric parameter $\mu$ all of the same order of magnitude,  which looks rather embarrassing in the context of the generic SUSY models. However, in the framework of the $SU(6)$ model \cite{Berezhiani:1989bd} with pseudo-Goldstone solution for the gauge hierarchy and Doublet-Triplet splitting problems this situation arises rather naturally. In this case in fact, the supersymmetric $\mu$-term emerges as a result of SUSY breaking, and 
 at the leading order approximation the condition of Eq. \eqref{eq:Higgs_finetunig} is straightforwardly obtained. 
 
In view of this, if one extends the needed conspiracy between the soft $F$- and $D$-terms in the Higgs sector to the squark and slepton masses ($D$-terms) and the gaugino masses ($F$-terms) as well, then all supersymmetric partners will be expected to have masses of the same order of magnitude.   

Let us remark also that for the proper condensation of the Higgs field a large quartic coupling constant ruling its self-interaction is needed, namely, one should have $\lambda = M^2_h/2v^2 \approx 0.13$.  On the other hand, in the framework of the MSSM such a coupling is mainly provided by the gauge $D$-terms of $SU(2)\times U(1)$. Hence, at a scale of the order of the mass terms present in Eq. \eqref{eq:Hmatrix} one has $\lambda = \frac14 (g^2 + g^{\prime^2}) \cos^2 2\beta$, where $g$ and $g'$ are the running gauge couplings  of $SU(2)$ and $U(1)$ respectively. At the electroweak scale the gauge couplings are  too small for saturating the needed value and one only gets $\lambda \approx 0.07$. From this result it is derived the famous Higgs mass limit in the MSSM $M_h^2 = M_Z^2 \cos^2 2\beta \,+$ radiative corrections, which makes difficult to recover the LHC Higgs mass $M_h \approx 125$~GeV and requires huge radiative corrections.  However, the running of the Standard Model couplings shows that $\lambda$ fastly decreases with energy \cite{Degrassi:2012ry}, and  at a scale of the order of 10 TeV it almost matches the needed value, unless $\tan\beta$ is too small. As it was shown in Ref. \cite{Arbey:2011ab}, by taking into account radiative corrections of reasonable size, the MSSM Higgs picture in the above described decoupling limit is indeed compatible with the 125 GeV Standard light Higgs. This is the case for a scale of SUSY larger then few TeV, and if $\tan\beta$ is not too small,  namely $\tan\beta \gtrsim 2$. 

Therefore, the Higgs physics, modulo specific conspiracies, indicates that the SUSY scale should be above few TeV, and similarly the bounds from the electric dipole moments settle the lower limit to few TeV as well. Generally flavor violations fix more stringent bounds, and for arbitrary  soft masses and $A$ parameters they would require SUSY scale larger than $\sim$ 100 TeV. However, there can be (flavor) symmetry reasons that providing approximate alignment of the soft SUSY parameters with the Yukawa couplings reduce the lower limit to $\sim$ 10 TeV. Moreover, in some theoretical scenarios based on flavor symmetries  the flavor limits can allow SUSY scales as small as 1 TeV  providing interesting relations between the soft mass matrices and SSB trilinear terms with the fermion Yukawa matrices as e.g.  $\tilde{m}^2_{eij}  = m_0^2 \delta_{ij} + m_1^2 (Y^{e \dagger} Y^e)_{ij} + m_2^2  (Y^{e \dagger} Y^e)_{ij}^2$ and $A^e_{ij} = m_3 Y^e_{ij}$, $m_{0,1,2,3}$ being soft mass parameters \cite{Berezhiani:1996ii,Berezhiani:1996kk,Anselm:1996jm,Berezhiani:2001mh}. Therefore, one can conclude that SUSY scale larger than few TeV is quite natural and if SUSY indeed lives near this lower bound, then it should be discovered in the LHC run II. 

\subsection{A generic SUSY-GUT in the SU(5) bottleneck}

In minimal SUSY $SU(5)$ \cite{Dimopoulos:1981zb, Sakai:1981gr}, the Higgs sector contains the following chiral superfields:  $\Sigma$ in the adjoint representation $\bf{24}$ of $SU(5)$ and  $H$, $\overline{H}$ respectively  in the fundamental representations $\bf{5}$, $\bf{\overline{5}}$. In this case one has the following expression for the Higgs part of the superpotential in Eq. \eqref{eq:superpotential}:
\begin{equation}
\label{eq:Higgs_superpotential}
W_{\rm Higgs} =
\frac{M_\Sigma}{2} \, \Sigma^2+\frac{\lambda_\Sigma}{3}\, \Sigma^3 + M_H \, H \overline{H}  + \xi \,  H \Sigma \overline{H}\,. 
\end{equation}
There exists a supersymmetric minimum in which the adjoint field $\Sigma$ gets a v.e.v.  breaking $SU(5)$ down to $SU(3) \times SU(2) \times U(1)$: 
\begin{equation}
\label{Sigma-VEV}
\langle \Sigma \rangle = V_\Sigma \times {\rm diag}\left(\,\frac23,\frac23,\frac23,-1,-1\right) \,, 
\quad\quad  V_\Sigma = \frac{3M_\Sigma}{\lambda_\Sigma} \,.
\end{equation}
As a result, the ``leptoquark"  gauge bosons $X,Y$ of $SU(5)$ get a mass
\begin{equation}
M_{\rm GUT} = M_{\rm X,Y} = \frac{5}{3}  \sqrt{2 \pi \alpha_{\rm GUT}} V_{\Sigma}\,, 
\label{eq:alpha_GUT}
\end{equation}
where $\alpha_{\rm GUT}$ is the $SU(5)$ gauge coupling, while the Higgs supermultiplets in the broken phase have masses
\begin{equation}
 \tilde{M}_{\Sigma} = \frac{5}{3} \lambda_\Sigma V_{\Sigma} \,, \quad
 \tilde{M}_D= \mu = M_H - \xi V_\Sigma \ll V_\Sigma \,, \quad 
  \tilde{M}_T = M_H + \frac23 \xi V_\Sigma \approx \frac53 \xi V_\Sigma \,, 
 \label{eq:lambda_GUT}
\end{equation}
In the previous expression $\tilde{M}_\Sigma$ denotes the mass of the color octet and weak isospin triplet fragments of the $SU(5)$ adjoint Higgs $\Sigma$ ($({\bf 8},{\bf1}) \oplus ({\bf1},{\bf 3})$ under $SU(3) \times SU(2)$), which are degenerate in mass. The quantity $\tilde{M}_T$ denotes the mass of the Higgs triplet supermultiplets $T,\bar T$ contained in $H,\bar H$, and $\tilde{M}_D$ is the mass ($\mu$-term in Eq. \eqref{eq:W-MSSM}) of their doublet fragments $H_u$ and $H_d$. The latter, which determines the Higgsino masses in the MSSM, should be small, while $\tilde{M}_T$ should be large, order the GUT scale, in order to avoid a too fast proton decay mediated by the color triplet Higgs or Higgsino exchanges. In minimal $SU(5)$ \cite{Dimopoulos:1981zb, Sakai:1981gr}, the price for such Doublet-Triplet (D-T) splitting is the {\it fine tuning} between the parameters in the superpotential  of Eq. \eqref{eq:Higgs_superpotential}. In particular, Eq. \eqref{eq:lambda_GUT} shows that two large values $M_H$, $\xi V_\Sigma\sim 10^{16}$ GeV should be fine tuned  for obtaining $\mu \sim 10^3$ GeV, with the precision of about $10^{-13}$. However, this condition is sufficient when one takes into account the SSB terms in Eq. (\ref{eq:ssb-lagrangian}), in which the function $\eta W'$  in this case can be presented as 
\begin{equation}
\label{eq:Higgs_Wprime}
\eta W'_{\rm Higgs} = \theta^2 \left(
\frac{M_\Sigma}{2} B_\Sigma \, \Sigma^2+\frac{\lambda_\Sigma}{3}\, A_\Sigma 
\Sigma^3 + M_H \, B_H H\overline{H}  +  \xi \, A_H H \Sigma \overline{H} \right) \,, 
\end{equation}
where $A$ and $B$ are dimensional parameters of the order of $M_F$. The presence of these terms shift the v.e.v.  of $\Sigma$ and generate its non-zero $F$-term, and at the end, the full account of these contributions generates the soft term $\mu B_\mu$ in Eq. \eqref{eq:F-MSSM}, as $\mu B_\mu = \xi V_\Sigma (A_H - A_\Sigma - B_H + B_\Sigma)$ + $\cal{O}$$(M_F^2)$. Therefore,  for adjusting this value ${\cal O}$$(V_\Sigma M_F)$ to the needed value $\mu B_\mu \sim M_F^2\sim 1~{\rm TeV}^2 $, the parameter $A_H - A_\Sigma - B_H + B_\Sigma$, which is ${\cal O}$$(M_F)$, should be extremely fine tuned. Therefore, this situation cannot be called a natural solution of the hierarchy and D-T splitting problems. 

In minimal $SU(5)$ model the problem of D-T splitting has only a technical solution, given by fine tuning of $\mu$ in Eq.~\eqref{eq:lambda_GUT} that is stable against radiative corrections. This situation is then worsened by the need of another fine tuning in the soft term $\mu B_\mu$. Of course, minimal $SU(5)$ model is not realistic, and one should not be surprised to find that fine tunings are required. 

Below we describe more natural situations when fine tunings can be avoided in specific GUT models. Nevertheless, we shall consider minimal $SU(5)$ as a prototype model, in the sense of $SU(5)$ bottleneck condition, for understanding the possible threshold corrections.\footnote{In particular, the masses $\tilde{M}_{\Sigma}$, $\tilde{M}_T$ are important for our analysis since, if they are smaller than $M_{\rm GUT}$, they would give significant threshold corrections near the GUT scale and affect the gauge coupling unification. However, within our paradigm of naturalness, couplings $\lambda_\Sigma$ and $\xi$ are assumed to be ${\cal O}$(1), and thus $\tilde{M}_{\Sigma},\tilde{M}_T \geq M_{\rm GUT}$. Nevertheless, we study also situation when these fragments can be relatively light, taking a possibility that within the natural spread of ${\cal O}$(1) values, some of the above couplings could be as small as $0.1$.}

There are several realistic models in which the D-T splitting problem can be solved without fine tunings. In particular, in $SU(5)$ this can be done via the ``Missing Doublet Mechanism" (MDM) \cite{Georgi:1981vf, Masiero:1982fe, Grinstein:1982um, Hisano:1994fn, Berezhiani:1996nu}, in $SO(10)$ via the ``Missing v.e.v.  Mechanism" (MVM) \cite{Dimopoulos:1981xm, Srednicki:1982aj, Babu:1994dq, Berezhiani:1996bv}, while in $SU(6)$ via the ``pseudo-Goldstones instead of Fine Tuning" (GIFT) Mechanism \cite{Inoue:1985cw, Anselm:1988ss, Berezhiani:1989bd, Barbieri:1993wz, Berezhiani:1995sb, Barbieri:1994kw}. In these models, the required patterns of the superpotentials are usually obtained by imposing some discrete symmetries, which are guaranteed at the level of renormalizable couplings, but  cannot provide suppression of dangerous high order operators destabilizing these solutions. It is interesting to note, however,  that stable solutions at any order can be achieved by making use of the anomalous $U(1)_A$ gauge symmetry (see Refs. \cite{Berezhiani:1996nu, Berezhiani:1996bv, Dvali:1996sr} respectively for the MDM $SU(5)$, MVM $SO(10)$ and GIFT $SU(6)$). In these models the soft parameters like $B_\Sigma$ and $A_\Sigma$ for the heavy GUT breaking superfields can be quite large, without creating additional fine-tuning problems.  Let us briefly describe them here.\\

In SU(5), the MDM \cite{Georgi:1981vf, Masiero:1982fe} contains the Higgs superfields in representations $\Phi \sim {\bf 75}$, $H\sim {\bf 5}$, $\bar H \sim {\bf \bar 5}$, $\Psi \sim {\bf 50}$, $\bar \Psi \sim {\bf \overline{50}}$, with the following superpotential terms:
\begin{equation}
W = M\Phi^2 + \lambda \Phi^3 + M_1 \Psi \bar{\Psi} + \lambda_1 H \Phi \bar{\Psi} + 
\lambda_2 \bar{H} \Phi \Psi + \mu H\bar{H} \,,
\end{equation}
with $M$ and $M_1$ being the mass parameters order $M_{\rm GUT}$ and $\lambda$'s being ${\cal O}$(1) coupling constants. $SU(5)$ is broken down to $SU(3)\times SU(2) \times U(1)$ by the v.e.v.  of $\Phi\sim {\bf 75}$. The latter represents a less economic multiplet replacing the adjoint $\Sigma$ of the minimal $SU(5)$, which also generates the mixing between the color triplet fragments in the Higgs 5- and 50-plets, whereas there are no doublets in the 50-plets. In this way, all color triplets are heavy, with mass of order $M_{\rm GUT}$, while the doublets in $H,\bar H$ remain light, with mass given by the $\mu$-term. In this theory the new soft parameter $B_\Phi$ (see Eq. \eqref{eq:Higgs_Wprime} for its analogous definition) can be taken large without inducing a large $B_\mu$. The unpleasant feature of this model is that the parameter $\mu$ should be taken very small with respect other masses $M,M_1$ in an {\it ad hoc} way.

Seemingly this model should satisfy our $SU(5)$ bottleneck condition, since it is based just on $SU(5)$. However, the situation is not so simple. The problem is that it involves the huge superfields ${\bf 75}$, ${\bf 50}$ and ${\bf \overline{50}}$, and if their fragments remain just slightly lighter than $M_{\rm GUT}$, their large threshold corrections can  completely ruin the gauge coupling unification. Moreover, the symmetry motivated versions of the MDM model \cite{Hisano:1994fn, Berezhiani:1996nu} contain twice as large amount of these fields. Therefore, in the context of these models the crossing of gauge couplings, with the todays precision, can be hardly considered stable for large threshold corrections. \\

For SO(10), in the MVM \cite{Dimopoulos:1981xm, Srednicki:1982aj}, the philosophy is similar: the Higgs doublets remain massless since some of the GUT-breaking fields have vanishing v.e.v.  along the direction that would give them a mass, whereas they couple with the triplets with non-zero v.e.v.. Therefore also in this case the protection of the doublet sector is due to group theoretical reasons. Therefore large soft terms $\sim10$~TeV in the heavy sector will not influence $\mu$ and $B_\mu$.\\
Regarding one step unification condition, it is not {\it a priori} guaranteed in $SO(10)$. In fact this gauge group can be broken down to $SU(3)\times SU(2) \times U(1)$ not passing through the $SU(5)$ bottleneck, but e.g. via Pati-Salam subgroup $SU(4) \times SU(2) \times SU(2)$ and subsequent symmetry breaking chains. This can be settled by {\it ad hoc} choice assuming that the Higgs superfields, which break $SO(10)$ down to $SU(5)$ subgroup, have the largest v.e.v.. However, there remains the problem that this model, and $SO(10)$ models in general, are not very economic as far as the Higgs sector is concerned. The minimal set of GUT superfields necessary to achieve the correct symmetry breaking and MVM pattern within the renormalizable superpotential, contains one 54-plet, at least three 45-plets, ${\bf 16} \oplus {\bf \overline{16}}$ and two 10-plets \cite{Babu:1994dq, Berezhiani:1996bv}, whereas in some versions the much larger multiples as ${\bf 210}$ and ${\bf 126} \oplus {\bf \overline{126}}$ are used. Large threshold corrections due to these fragments can easily affect the crossing of gauge couplings and thus transform the gauge coupling unification from a prediction to an accidental fact.\\ 

The $SU(6)$ model \cite{Berezhiani:1989bd, Barbieri:1993wz, Berezhiani:1995sb, Barbieri:1994kw,Berezhiani:1995dt,Dvali:1996sr} is based on very simple set of GUT superfields, in fact as simple as the minimal $SU(5)$, $SU(6)$ gauge symmetry is broken by two sets of superfields:  two fundamental representations $H\sim {\bf 6}$ and $\bar H \sim {\bf \overline 6}$ that break $SU(6)$ down to $SU(5)$, and one in adjoint representation $\Sigma \sim {\bf 35}$, that leads to the breaking channel $SU(6) \to SU(4) \times SU(2) \times U(1)$. As a result the two channels together break the $SU(6)$ gauge symmetry down to $SU(3) \times SU(2) \times U(1)$. The simple assumption on which the GIFT mechanism is based is that the Higgs superpotential does not contain the mixed term $H \Sigma \bar H$, so that it has the form $W = W(\Sigma) + W(H,\bar H)$, where
\begin{equation}
W(\Sigma) = \frac{M_\Sigma}{2}  \Sigma^2 + \frac{\lambda_\Sigma}{3} \Sigma^3\,,
\qquad W(H,\bar H) = Y (H \bar H - V^2)\,.
\end{equation}
As a result the superpotential acquires an accidental global symmetry $SU(6)_\Sigma \times SU(6) _H$, which independently transforms $\Sigma$, and $H$, $\bar H$ superfields. Then, in the limit of unbroken supersymmetry the MSSM Higgs doublets $H_u$, $H_d$ appear as massless Goldstone superfields built up as a combination of doublet fragments from $\Sigma$, and $H$, $\bar H$, that remain uneaten by the gauge bosons. Therefore in this limit $\mu$ vanishes exactly.

Supersymmetry breaking terms like $A_\Sigma$, $B_\Sigma$ shift the v.e.v.'s and also give F-terms to them, therefore generating $B_\mu$ term for the MSSM Higgses. However, since these terms also respect the global symmetry $SU(6)_\Sigma \times SU(6)_H$, the mass matrix of the Higgses in Eq.~\eqref{eq:Hmatrix} is degenerate, and thus one Higgs scalar (combination of the scalar components of $H_u$ and $H_d$) still remains massless.  Thus, even with arbitrary $B_\Sigma$ that give $\mu\sim B_\mu\sim B_\Sigma$, there is an automatic relation between $\mu$ and $B_\mu$ terms that guarantees that the determinant of Eq. \eqref{eq:Hmatrix} vanishes. This degeneracy is removed only by radiative corrections due to Yukawa terms that do not respect the global symmetry, and the resulting Higgs mass will be of the order of $\mu$ and $B_\mu$, given by the mismatch in their renormalization. Therefore, in the case of large $B_\Sigma \sim 10$~TeV we are still left with a ''little'' hierarchy problem of the electroweak scale stability against $10$~TeV scale discussed in previous section. 

Due to economic Higgs sector and the way in which the GIFT mechanism works in $SU(6)$, the $SU(5)$ bottleneck condition is satisfied in straightforward way. In a most elegant way this occurs in the context of model \cite{Dvali:1996sr} with anomalous $U(1)_A$ symmetry where $H,\bar H$ v.e.v.'s emerge essentially  at the string scale. Below such energy the  theory becomes a minimal $SU(5)$ with one adjoint ${\bf 24}$ and two fundamentals ${\bf 5}$ and ${\bf \overline 5}$ plus one $SU(5)$-singlet, altogether composing a 35-plet of $SU(6)$ that remains a global symmetry at this point.\\ 

Until now we have not considered yet the fermion sector and the Yukawa part of the superpotential in Eq. \eqref{eq:superpotential}. Coming back to a naive minimal $SU(5)$ where fermions are allocated in the representations ${\bf 10}_i \oplus {\bf \overline{5}}_i$, $i=1,2,3$, one can write the Yukawa couplings 
\begin{equation}
\label{eq:W-SU5}
W_{\rm Yukawa} = 
Y^u _{ij} {\bf 10}_i {\bf 10}_j H  + Y^e _{ij} {\bf 10}_i {\bf \overline 5}_j \bar H \,,
\end{equation}
which after $SU(5)$ breaking reduces to the Yukawa part of $W_{\rm MSSM}$ reported in Eq. \eqref{eq:W-MSSM}. However, one immediately encounters two problems:\\
i) this predicts $Y^d_{ij} = Y^e_{ij}$ and thus degeneracy of the Yukawa eigenvalues between down quarks and leptons of all generations at the GUT scale, $y_{d,s,b} = y_{e,\mu,\tau}$. While for the third generation the $y_b = y_\tau$ unification works (almost) perfectly, for the lighter generations it is dramatically wrong;\\
ii) in this way we must introduce small Yukawa couplings for lighter generations. In fact while only top quark has a large Yukawa $y_t \simeq 1$ (for  large $\tan\beta$, $y_b$ and $y_\tau$ could be large as well), the rest of the Yukawa couplings must be $\ll 1$ in any case. This would contradict our assumption of no artificially small parameters in the mother GUT theory. 

Fortunately, both these problems can be solved at one shot in a rather natural manner, assuming that, by some symmetry reasons\footnote{Typically due to family symmetries (in $SU(6)$ model this occurs automatically due to pseudo-Goldstone nature of the Higgs superfields) only top quark can get the mass from renomalizable coupling with $y_t\sim 1$, while other masses come from higher order terms suppressed by $M_P$. It is worth to notice, however, that in spite masses of $b$ and $\tau$ emerge from higher order operator, $y_b = y_\tau$ still holds with the modulo corrections related to hierarchy between the $SU(6) \to SU(5)$ and $SU(5) \to SU(3)\times SU(2) \times U(1)$ breaking scales  with $V_5/V_6 \sim 10^{-1} \div10{^{-2}}$ \cite{Barbieri:1994kw, Berezhiani:1995dt}.}, only the third generation gets masses from the renormalizable Yukawa terms of Eq. \eqref{eq:W-SU5}, while the masses of the first two generations are due to higher order operators 
\begin{equation}
\label{eq:yu}
 a^u_{ij} {\bf 10}_{i} {\bf 10}_{j} \frac{\Sigma H}{M_{P}} + b^u_{i j} {\bf 10}_i {\bf 10}_j \frac{\Sigma^{2} H}{M^{2}_{P}}
+a^d_{ij} {\bf 10}_{i} {\bf \overline{5}}_{j} \frac{\Sigma \overline{H}}{M_{P}} 
+b^d_{ij} {\bf 10}_{i} {\bf \overline{5}}_{j} \frac{\Sigma^2 \overline{H}}{M_{P}^2} + .... 
\end{equation}
In the previous expression $M_P$ denotes the Planck energy scale and $i,j=1,2,3$, with the coefficients $a,b$ $\sim \cal{O}$(1) (see e.g. in \cite{Berezhiani:1995tr}). The expression \eqref{eq:yu} after $SU(5)$ breaking gives the Yukawa terms for the light families and their mixing with the third family. In this case we gain two things: the light fermion Yukawa couplings are no more degenerate, since they get Clebsch factors from the v.e.v.  of $\Sigma$, and hierarchy between the families can be naturally understood in terms of small factor $\langle \Sigma \rangle/M_P \sim 10^{-1} \div 10^{-2}$. The effective high order operators can be obtained by integrating out some heavy fermions in vector-like representations ${\bf 5} \oplus  {\bf \overline 5}$, ${\bf 10} \oplus {\bf \overline{10}}$, with masses ${\cal O}(M_P)$, with which light chiral families ${\bf 10}_i \oplus {\bf \overline{5}}_i$ mix with the $\cal{O}$(1) Yukawa couplings \cite{Froggatt:1978nt, Berezhiani:1983hm, Berezhiani:1985in, Dimopoulos:1983rz}. In this way, the mother theory becomes free of small couplings, namely all coupling constants are $\cal{O}$(1). It should be also noticed that due to these terms, the $y_b=y_\tau$ unification in third family is not anymore exact but it gets corrections, typically $y_b=y_\tau \big (1 \pm {\cal O}(y_\mu/y_\tau)\big)$ while in some predictive scenarios with asymmetric textures \cite{Berezhiani:1996bv, Berezhiani:2000cg, Berezhiani:1998vn} corrections could be even larger. 

This situation emerges in a very natural way in the context of $SU(6)$ model \cite{Berezhiani:1989bd}. Its minimal anomaly free fermion content contains three generations of chiral superfields in multiplets that are a $SU(6)$ decomposition of  a {\bf 27}-plet of $E_6$, ${\bf 27} = \bar{\bf 6} \oplus \bar{\bf 6} \oplus {\bf 15}$, containing two (anti)fundamental representations, and a two-index antisymmetric one.\footnote{For a comparison, $SO(10)$ decomposition of the $E_6$ {\bf 27}-plet is $ {\bf 16} \oplus {\bf 10} \oplus {\bf 1}$ containing the spinor representation {\bf 16}, but also additional non chiral fragments ${\bf 10} \oplus {\bf 1}$.} Their $SU(5)$ decomposition reads $\bar{\bf 6} = \bar{\bf 5} \oplus \bar{\bf 1}$ and ${\bf 15} = {\bf 5} \oplus {\bf 10}$. The only possible Yukawa couplings are $g_{ik}  {\bf 15}_i \bar{\bf 6}_{k} \bar{H}$, $i=1,2,3$ and $k=1,2,...6$, where the matrix $g_{ik}$ can be taken block diagonal without lose of generality, e.g. with non-vanishing elements being only $g_{14},g_{25},g_{36}$. After symmetry breaking $SU(6) \to SU(5)$ by the v.e.v. of $\bar{H}$ ($V_H$),  these couplings combine  three fragments ${\bf 5}_i \subset {\bf 15}_i$ with  three combinations of six $\bar{\bf 5}$ fragments from  $\bar{\bf 6}_k$ into massive particles with $M\sim V_H$. Consequently, theory reduces to $SU(5)$ with three generations ${\bf 10}_i \oplus \bar{\bf 5}_i$, $i=1,2,3$.  However, non-renormalizable Yukawa couplings can be written that could generate fermion masses. 

It is possible to make a minimal modification of the model by introducing a superfield ${\bf 20}$  (three-index antisymmetric representation) \cite{Barbieri:1994kw,Berezhiani:1995dt} whose  $SU(5)$ decomposition is ${\bf 20} = {\bf 10} \oplus \overline{\bf 10}$. Since ${\bf 20}$ is a pseudo-real representation of $SU(6)$, its mass term is not allowed: convolution of ${\bf 20}^2$ with a Levi-Chivita tensor is vanishing. Therefore, its ${\bf 10} \oplus \overline{\bf 10}$ fragments remain massless before $SU(6)$ symmetry breaking. In this case, the Yukawa couplings $\lambda \, {\bf 20} \, {\bf 20} \, \Sigma$ and $\lambda'_i \, {\bf 20} \, {\bf 15}_i \, H$ that explicitly violate the global $SU(6)_\Sigma \times SU(6)_H$ symmetry can be introduced. The latter one combining $\overline{\bf 10} \subset {\bf 20}$ with ${\bf 10} \subset {\bf 15}$ provides massive states, while the former one reduces to a $SU(5)$ Yukawa coupling for the fragment ${\bf 10} \subset {\bf 20}$ with ${\bf 5}$ fragment of $\Sigma$, $\lambda \, {\bf 10} \, {\bf 10} \,  {\bf 5}_\Sigma$. In this way, an upper quark from ${\bf 10}$, to be identified with top quark, gets Yukawa coupling with the pseudo-Goldstone fragment $H_u \subset  {\bf 5}_\Sigma$, with $y_t \sim \lambda \sim 1$. The $t$-quark is the only particle getting in this way  mass ${\cal{O}}$(100) GeV \cite{Barbieri:1994kw,Berezhiani:1995dt}. Masses of other particles should necessarily emerge from the higher order operators.  Namely, masses of $b$ and $\tau$ come from the operator $(\lambda_b/M_P^2) \, {\bf 20} \, \bar{\bf 6}_3 \bar{H}^2 \Sigma$ implying $y_b = y_\tau$ modulo small corrections, namely $y_b = y_\tau \big (1 + {\cal O}(V_\Sigma /V_H)\big)$. The mass of $c$-quark is induced by the term $(\lambda_c/M_P^2) \, {\bf 15}_2 {\bf 15}_2 {H}^2 \Sigma$, with a nice implication $y_c \sim y_{b,\tau} \sim (V_H/M_P)^2$ pointing towards small or moderate $\tan\beta$. It is worth to remark that this feature also implies the lower bound $(V_H/M_P)^2 > 0.01$, i.e. $V_H \gtrsim 10^{17}$~GeV. Hence, $V_H \gg V_\Sigma$ and the $SU(5)$ bottleneck condition is naturally satisfied. 

In this framework, the SSB D-terms of Eq. \eqref{eq:W-SU5}, which generate masses for squarks and sleptons, are given by
\begin{equation}
\mathcal{L}_{\rm Y, D}^{\rm SU(5)}=\tilde{m}^{2}_{{\bf 10}ij} {\bf \tilde{10}}^{\dag}_i {\bf \tilde{10}}_j + \tilde{m}^{2}_{{\bf 5}ij} {\bf \tilde{5}}^{\dag}_i {\bf \tilde{5}}_j\,. \label{eq:su(5)_dterm}
\end{equation}
In our analysis, we allow the mass matrices $\tilde{m}_{\bf 10}$ and $\tilde{m}_{\bf 5}$ to be different. Different squark and slepton fragments, which have the same soft masses in the GUT limit, split at low energies due to the running. The effect of such difference on the compatibility of the model has been considered in the following analysis. Note that  $\tilde{m}_{\bf 10}$ and $\tilde{m}_{\bf 5}$ would be exactly equal in $SO(10)$ inspired models at  $M_{\rm GUT}$. As for inter-family splitting of soft masses, as well as for healthy pattern of the trilinear terms in Eq.~\eqref{eq:F-MSSM}, they can be alligned with the Yukawa terms through MFV relations between the Yukawa couplings and soft parameters for squarks and sleptons,  as e.g. $\tilde{m}_{uij}^2 = m_0^2 \delta_{ij} + m_1^2 (Y^{u\dagger} Y^u)_{ij} + m_2^2  (Y^{u \dagger} Y^u)^2_{ij}$ and $A^u_{ij} = m_3 Y^u_{ij}$, where $m_{0,1,2,3}$ are soft mass parameters. Therefore, due to small values of the Yukawa couplings, mass spectrum of the squarks and sleptons must have inter-family degeneracy with possible exclusion of {\it stop} since $y_t \sim 1$. In any case, this unification of fermion and sfermion masses and interaction patterns provides a chance for a natural suppression of SUSY induced flavor violating effects, and allows to lower the SUSY scale down to 1 TeV.  It is important that such relations can be obtained also in the context of realistic and predictive SUSY-GUT scenarios for the fermion masses and mixing \cite{Berezhiani:1996ii, Berezhiani:1996kk, Anselm:1996jm, Berezhiani:2001mh, Berezhiani:2005tp}.

\section{Renormalization Group Equations: initial conditions and threshold corrections}

Let us consider  the Renormalization Group Equations (RGEs), up to 2-loop order, for a set of couplings $X_i$ that can be generally cast in the following form
\begin{equation}
\frac{d}{dt}X_i=\frac{1}{16\pi^{2}}\beta_{X_i}^{(1)}(X_j)+\frac{1}{(16\pi^{2})^{2}}\beta_{X_i}^{(2)}(X_j)\,,
\end{equation}
where $t=\ln(M/M_{0})$, and $M_0$ represents the renormalization energy scale at which we impose the initial conditions. The set of running couplings and the corresponding expressions for $\beta_{X_i}^{(1)}$ and $\beta_{X_i}^{(2)}$ depend on the energy regime considered, simply denoted as SM (non SUSY) or MSSM (SUSY), and on the massless particle content at that particular energy scale. The complete set of SM RGEs is provided by \cite{Mihaila:2012fm, Mihaila:2012pz, Mihaila:2012bt, Luo:2002ey, Chetyrkin:2012rz, Bednyakov:2012en}, while the MSSM RGEs are reported in \cite{Martin:1993zk}. We properly assume that only the Yukawa couplings for the heaviest particles of each family generation (i.e. top and bottom quarks and tau lepton) give a considerable contribution. Moreover, in the running of gauge couplings we consider only the top Yukawa contribution since the others are negligible at all energy scales for moderate $\tan\beta$. Moreover, in the MSSM $\beta$-functions we neglect the terms related to trilinear soft couplings $A_{ij}^{u,d,e}$ of Eq.~\eqref{eq:F-MSSM}.

As usual, starting from low energy,  it is convenient to fix the renormalization mass $M_0$  at the EW scale $M_{\rm Z}=91.1876\pm0.0021$ GeV  \cite{Agashe:2014kda}, where the gauge couplings $\alpha_i=g^2_i/4\pi$ result to be  in the $\overline{MS}$ renormalization scheme \cite{Agashe:2014kda}
\begin{equation}
\alpha_1^{-1}(M_{\rm Z})=59.008\pm0.008\,,\hspace{7.mm}\alpha_2^{-1}(M_{\rm Z})=29.569\pm0.014\,,\hspace{7.mm}\alpha_3(M_{\rm Z})=0.1184\pm0.0007\,,
\end{equation}
where $i=1,\,2,\,3$ stands for $U(1)$, $SU(2)$, $SU(3)$, respectively. As it is well known, in the running from low to high energy scale we have to add every particle above its individual threshold given by its mass, and to impose at the same time corresponding matching conditions. In the SM this happens, for instance, in the case of Higgs  ($M_h=125.7\pm0.4$ GeV) and of top quark ($M_t=173.21\pm0.51\pm0.71$ GeV)  \cite{Agashe:2014kda}. 

Starting from $M_{\rm Z}$, with the increase of energy we have at a certain scale the transition from SM to MSSM. Here, we have two possibilities, namely, the {\it single-scale} and {\it multi-scale approaches} \cite{Baer:2005pv}. In the single-scale approach, the transition between SM and MSSM occurs at a given {\it effective} energy scale simply denoted as $M_{\rm SUSY}$. In this scenario all SUSY particles share the same mass. 
\begin{figure}[tbp]
\centering
\includegraphics[width=.40\textwidth]{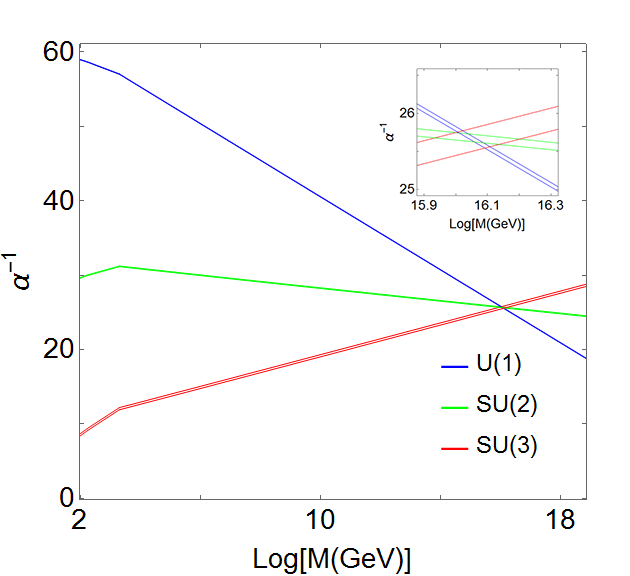}
\caption{\label{fig:RGE}Running of the gauge couplings $\alpha_1^{-1}$ (blue), $\alpha_2^{-1}$ (green) and $\alpha_3^{-1}$ (red) with $M_{\rm SUSY}=2$ TeV up to 2-loop in $\beta$-functions. For each $\alpha^{-1}$ it is reported the $3 \sigma$ error band due to the experimental uncertainty at $M_{\rm Z}$. The range for $M_{\rm SUSY}$ compatible with the measurements within $3\sigma$ is from $\sim1$ TeV up to $\sim3.5$ TeV. The inset is a focus of the crossing region.}
\end{figure}
In Fig.\ref{fig:RGE} one shows the gauge couplings unification in a SUSY scheme with a single effective energy scale $M_{\rm SUSY} \sim TeV$. In achieving such result, as it is well known, 2-loop $\beta$-functions play a relevant role in increasing the value for $M_{\rm SUSY}$ needed.  However, the simplifying assumption of single-scale approach is very unnatural, since, in principle, the SUSY particles mass spectrum might be not degenerate. When such occurrence is considered we have the so-called  multi-scale approach that is adopted in our analysis.

In the framework of SUSY-GUTs under the {\it SU(5) bottleneck} ansatz, sparticles, even though with the same mass at $M_{\rm GUT}$, naturally acquire different masses al low scale due to the renormalization group equations. In our model, according to Eqs. \eqref{eq:F-MSSM} and \eqref{eq:su(5)_dterm}, at $M_{\rm GUT}$ there exist three parameters related to SUSY soft masses $\tilde{m}_{\rm G}$ (SSB F-terms) for gauginos and, $\tilde{m}_{\bf 10}$ and $\tilde{m}_{\bf 5}$ (SSB D-terms) for matter sparticles.  Let us denote with $\tilde{m}_{\rm g}$, $\tilde{m}_{\rm W}$, $\tilde{m}_{\rm sq}$ and $\tilde{m}_{\rm sl}$ the pole masses of gluinos, neutralinos, squarks and sleptons at low energy scale, respectively. Due to the {\it SU(5) bottleneck} ansatz, at $M_{\rm GUT}$ we can assume $\tilde{m}_{\rm g}/ \tilde{m}_{\rm W} =1$ and  $\tilde{m}_{\rm sq}/\tilde{m}_{\rm sl }= \tilde{m}_{\bf 10}/\tilde{m}_{\bf 5} =1$. It is worth while noticing that the simplicity assumption $\tilde{m}_{\bf 10}/\tilde{m}_{\bf 5} =1$ does not sensibly affect the final results  since the main contribution to gauge coupling running comes from gauge and Higgs sectors as will be clarified in the following. Finally,  there exists also another parameter $\tilde{m}_{\rm h}$ that is the mass of two higgsinos of the same order of supersymmetric parameter $\mu$. Note that the mass of the second Higgs is of the same order of the masses of matter sparticles.

Let us define $M_{\rm min}$ and $M_{\rm max}$ as the minimum and the maximum of SUSY particles mass spectrum, respectively, in a given model. Concerning the SUSY thresholds we adopt the following simplicity ansatz. We use below $M_{\rm min}$ 2-loop SM RGEs, while above $M_{\rm max}$  2-loop MSSM RGEs. Within these two scales, we apply 1-loop MSSM RGEs for each new SUSY particle, i.e. adding the contributions of each particles at the corresponding mass (see Ref.~\cite{Jones:1981we}), while we use 2-loop SM RGEs for SM particles.

Moreover,  since the SM $\beta$-functions are typically evaluated in the $\overline{MS}$ renormalization scheme, while the MSSM ones are obtained using the $\overline{DR}$, one should take into account matching relations between the couplings corresponding to  the different schemes. However,  since the transition between $\overline{MS}$ and $\overline{DR}$ gives a shift in the unknown SUSY particles masses smaller than 1\%, for the level of accuracy of the present analysis we can safely neglect such effect.

For Yukawa couplings of the corresponding particles, we use 2-loop SM RGEs up to $M_{\rm max}$ and 2-loop MSSM RGEs above such scale, imposing the suitable matching conditions 
\begin{equation}
y_t^{SM}=y_t^{MSSM}\sin\beta\,, \hspace{7.mm}
y_b^{SM}=y_b^{MSSM}\cos\beta\,, \hspace{7.mm}
y_\tau^{SM}=y_\tau^{MSSM}\cos\beta\,.
\label{eq:yukawa}
\end{equation}
Finally, GUT particles which are lighter than leptoquarks ($X$, $Y$) modify RGEs before gauge couplings unify at $M_{\rm GUT}$. For instance, the mass of the $({\bf 8},{\bf1}) \oplus ({\bf1},{\bf 3})$  fragments of $SU(5)$ adjoint Higgs,  $\tilde{M}_\Sigma$,  or of triplet fragments $({\bf 3},{\bf 1}) \oplus ({\bf \bar3},{\bf1})$ in $H,\bar H$,  could be  below $M_{\rm GUT}$ if the coupling constants $\lambda_\Sigma$ or $\xi$  in Eq.~\eqref{eq:Higgs_superpotential} are  small. Indeed, by comparing Eqs.~\eqref{eq:alpha_GUT} and \eqref{eq:lambda_GUT} we can define the following parameters 
\begin{equation}
 \chi_\Sigma \equiv \frac{M_{\rm GUT}}{\tilde{M}_\Sigma} = \frac{\sqrt {2 \pi \alpha_{\rm GUT}}}{\lambda_\Sigma}\, , 
 \quad 
 \chi_T \equiv \frac{M_{\rm GUT}}{\tilde{M}_T} = \frac{\sqrt {2 \pi \alpha_{\rm GUT}}}{\xi}\, ,
\label{eq:chi_GUT}
\end{equation}
If the parameter $\chi_\Sigma$ is large, then the GUT threshold corrections due to fragments of $\Sigma$  will be relevant. According to our naturalness conditions, we assume that $\lambda_\Sigma$ is not unnaturally small, and impose the constraint $\chi_\Sigma < 10$. The same can be applied also for color triplet fragments $T,\bar T$. However, the proton stability against dimension-5 operators require $\tilde{M}_T > M_{\rm GUT}$.  Therefore, $ \xi \sim 1$ is favored, and threshold corrections related to $\chi_T$ are expected to be irrelevant. It is interesting to notice that in the context of $SU(6)$ model \cite{Berezhiani:1989bd},  one automatically gets $\tilde{M}_\Sigma = \tilde{M}_T$ as far as they emerge from the same  adjoint Higgs $\Sigma \sim {\bf 35}$.  Hence, the requirement of the proton stability against Higgsino mediated dimension-5 operators, $\tilde{M}_T \geq M_{\rm GUT}$,  also implies that there should be no threshold corrections neither from other fragments the superfield $\Sigma$. 
  
The $b$-$\tau$ unification at GUT energy scale is also studied, by imposing at $M_{\rm GUT}$ 
\begin{equation}
y_{\rm b}(M_{\rm GUT}) =y_{\rm \tau}(M_{\rm GUT}) \left(1 + {\cal O}\left( \frac{y_{\rm \mu}(M_{\rm GUT})}{y_{\rm \tau}(M_{\rm GUT})}\right)\right)\,.
\label{eq:btau}
\end{equation}
Note that since the ratio between two Yukawa couplings is essentially unchanged by the running, we have used their values measured at EW scale in order to estimate ${\cal O}\left( y_{\rm \mu}/y_{\rm \tau}\right)$. The masses are $m_{\rm b}(m_{\rm b})=4.18 \pm 0.03$~GeV (evaluated in $\overline{MS}$ scheme), $M_{\rm \tau}=1776.82 \pm 0.16$~MeV and $M_{\rm \mu}=105.6583715 \pm 0.0000035$~MeV \cite{Agashe:2014kda}.

We have developed a \textit{Mathematica} program\footnote{Other programs able for such a purpose are for instance RunDec \cite{Chetyrkin:2000yt}, SPheno \cite{Porod:2003um} and SoftSusy \cite{Allanach:2001kg}.} which resolves all the RGEs with numerical iterative method and takes into account all the matching and threshold relations discussed in the previous sections.
\begin{figure}[tbp]
\centering
\includegraphics[width=.45\textwidth]{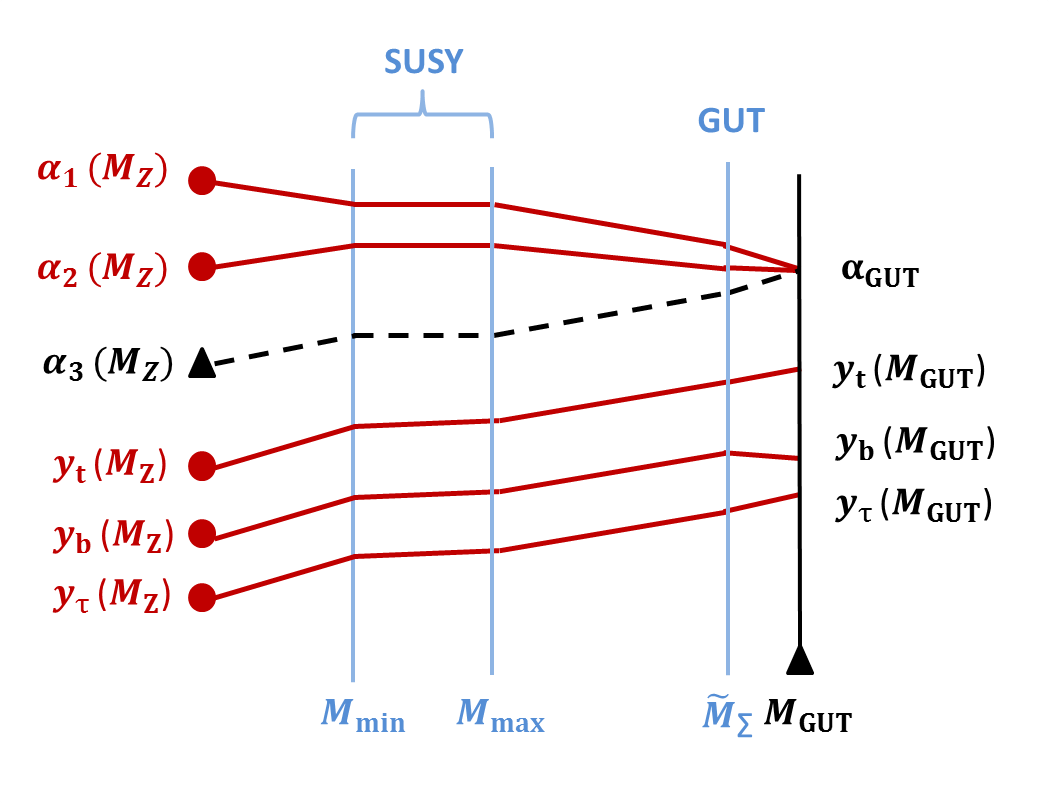}
\caption{\label{fig:program}Steps of the iterative method implemented in the \textit{Mathematica} program which solves the RGEs, imposing gauge unification at $M_{\rm GUT}$. All red solid lines mean that the differential equations are resolved from $M_{\rm Z}$ up to $M_{\rm GUT}$, whereas the black dashed line means that the corresponding RGE is resolved from high to low energy scales. The circles represent the initial conditions imposed at $M_{\rm Z}$, while the black triangles denote the output parameters that must be compatible with experimental bounds. The energy scales, $M_{\rm min}$ and $M_{\rm max}$, are respectively the minimum and the maximum of SUSY particles masses in a given model. The quantity $\tilde{M}_\Sigma$ denotes the mass of $\Sigma$ Higgs fragments $\left({\bf 8},{\bf 1}\right) \oplus \left({\bf 1},{\bf 3}\right)$. Note that all these energy scales define the SUSY and GUT thresholds.}
\end{figure}
The input parameters of the program are $\{\tilde{m}_{\rm h},\,\tilde{m}_{\rm g},\,\tilde{m}_{\rm sq},\,\chi_\Sigma,\,\tan\beta\}$ (hereafter simply denoted as a {\it model}), whereas the outputs are {\small{$ \{\alpha_3(M_{\rm Z}),\,M_{\rm GUT},\,\alpha_{\rm GUT},\,y_{\rm t}(M_{\rm GUT}), \,y_{\rm b}(M_{\rm GUT}), \,y_{\rm \tau}(M_{\rm GUT})\}$}}. The quantity $\chi_T$ has not been considered in the analysis since it contributes in an almost irrelevant way $(\xi \sim 1)$ to the running.

Since the output quantity  $\alpha_3(M_{\rm Z})$ must be compatible with its experimental value it represents a compatibility constraint that has to be satisfied by the particular model chosen. In other words, scanning on the possible input values one has to discard those choices producing a value of $\alpha_3(M_{\rm Z})$ outside the experimental range. This yields, at the end of the procedure, a set of compatible SUSY-GUT models.

Another compatibility constraint concerns $M_{\rm GUT}$ since it is straightforwardly related to the proton lifetime $\tau_p$. In a generic SUSY-GUT scheme, the proton decay is induced by dimension-6 operators (mediated by gauge bosons $X,Y$), and by any type of effective operators allowed by a general operator analysis \cite{Weinberg:1979sa, Wilczek:1979hc}. In this study we focus our attention on dimension-6 operators  responsible for the  relevant decay channel $p \rightarrow e^+\pi^0$, namely the exclusive decay channel posing the most stringent bounds. The corresponding experimental bound is $\Gamma^{-1}\left( p \rightarrow e^+\pi^0 \right) = \tau_p/{\rm Br}\left( p \rightarrow e^+\pi^0 \right)> 1.29 \cdot 10^{34}$ yr \cite{Nishino:2012ipa} with 219.7 kt-yr of data at 90\% confidence level. From Ref. \cite{Hisano:2012wq} one gets for the partial width
\begin{equation}
\Gamma\left( p \rightarrow e^+\pi^0 \right)  = \frac{\pi}{4}\frac{\alpha^2_{\rm GUT}}{M^4_{\rm GUT}} \frac{m_p}{f^2_\pi}\alpha^2_H 
| 1 + D + F|^2 \left( 1 - \frac{m^2_\pi}{m^2_p} \right)^2 \left[ \left(A^{(1)}_{\rm R} \right) +  \left(A^{(2)}_{\rm R} \right) \left( 1+ \left| V_{ud}\right|^2 \right)^2 \right] \, .
\label{eq:exproton}
\end{equation}
By substituting in the above expression: $f_\pi = 0.13$ GeV, $\alpha_H = -0.0112$ GeV$^3$, $D=0.8$, $F=0.47$, $A^{(1)}_{\rm R} \approx 2.5$ and $A^{(2)}_{\rm R} \approx 2.6$ (see Ref. \cite{Hisano:2012wq} for the definition of the parameters) and using \cite{Agashe:2014kda} one gets
{\small
\begin{equation}
\Gamma^{-1}\left( p \rightarrow e^+\pi^0 \right) \approx 2.4 \cdot 10^{32} \left( \frac{M_{\rm GUT}}{10^{16} \, \text{GeV}} \right)^4 \frac{1}{\alpha_{\rm GUT}^2}\,\text{yr} > 1.29 \cdot 10^{34} \text{yr} 
\hspace{3.mm} \implies \hspace{3.mm} \frac{M_{\rm GUT}}{\sqrt{\alpha_{\rm GUT}}} \gtrsim 3 \cdot 10^{16} \,\text{GeV}\,.
\label{eq:proton}
\end{equation}}
In the following we adopt this conservative limit that does not depend strongly on the details of the fermion masses and mixing. It is interesting to remark that the new effective operators of dimension-4 and -5 can be induced by the SSB terms when the heavy gauge bosons of GUT are integrated out \cite{Derendinger:1982tq, Berezhiani:2006mt}.  After being dressed by gauginos, they transform into dimension-6 operators, and in the case of large soft parameters, as it can be in our case, the proton lifetime in gauge mediated channels can be strongly affected (enhanced or even suppressed) depending on the pattern of soft terms in the heavy Higgs sector \cite{Berezhiani:2006mt}.   

The dimension-6 operators induced by the exchange of heavy color triplet scalars in $H,\bar H$ are suppressed by small Yukawa couplings of the first generation fermions, even though in some models with flavor symmetry \cite{Berezhiani:1984cg, Berezhiani:1983rk} their contribution can be dominant. In generic SUSY-GUTs,  the dominant contribution to proton decay usually comes from dimension-5 operators induced by the color triplet Higgsino exchange \cite{Weinberg:1981wj, Sakai:1981pk}.  After being dressed by gauginos, they induce proton decay dominantly via the channel $p \rightarrow K^+\overline{\nu}$. The experimental limit on the latter is quite stringent, namely $\tau_p/{\rm Br}\left( p \rightarrow K^+\overline{\nu}  \right) > 5.9 \cdot 10^{33}$ yr  at 90\% confidence level \cite{Abe:2014mwa}. This bound excludes the {\it naive}  supersymmetric $SU(5)$  model \cite{Goto:1998qg,Murayama:2001ur}, which is however already excluded by the wrong prediction for the light quark masses. For models reproducing realistic fermion masses the impact of these dimension-5 operators is strongly model dependent. In literature have been discussed many examples of models in which dimension-5 operators can be suppressed by particular symmetry reasons \cite{Hisano:1994fn, Hisano:1992ne, Babu:1993we, Berezhiani:1998hg, Berezhiani:1995yk, Dvali:1995hp}. By taking the constant $\xi \sim 1$, the color  triplet Higgsinos move to mass scales higher than the GUT scale. In addition, having  squark mass large enough, say 10 TeV,  the Higgsino mediated dimension-5 operators indeed become safe.  

As discussed before, the running of all couplings is evaluated using 2-loop $\beta$-functions, while the threshold effects are studied at 1-loop level only. Since the 2-loop RGEs are an extremely involved set of coupled equations, we employ an iterative method whose steps are shown in Fig.\ref{fig:program}. In order to impose the gauge unification at $M_{\rm GUT}$, for a given choice of input parameters and starting from the known values of $M_{\rm Z}$, $M_{\rm h}$, $M_{\rm t}$, $M_{\rm b}$, $M_{\rm \tau}$, $\alpha_1^{-1}(M_{\rm Z})$ and $\alpha_2^{-1}(M_{\rm Z})$ previously reported, we iteratively
\begin{itemize}
\item evolve $\alpha_{1}$ and $\alpha_{2}$, from $M_{\rm Z}$ to higher energy scales;
\item find their intersection point that defines the values of $\alpha_{\rm GUT}$ and its corresponding unification scale $M_{\rm GUT}$;
\item evolve backward $\alpha_3$ using $(\alpha_{\rm GUT}, M_{\rm GUT})$ as initial point, obtaining $\alpha_3(M_{\rm Z})$;
\item starting from this solution for running gauge coupling constants we introduce the contributions of $y_{\rm t}$, $y_{\rm b}$ and $y_{\rm \tau}$ in such a way that they consistently reproduce the known values of $M_{\rm t}$, $M_{\rm b}$ and $M_{\rm \tau}$. This fixes at the end the corresponding values for $y_{\rm t}(M_{\rm GUT})$, $y_{\rm b}(M_{\rm GUT})$ and $y_{\rm \tau}(M_{\rm GUT})$. Note that the other Yukawa couplings are not considered since they give subdominant contributions.
\end{itemize}
This procedure must be iterated until it converges. The convergence conditions is that the absolute difference $\left| \alpha^{i+1}_3(M_{\rm Z})-\alpha^i_3(M_{\rm Z}) \right| < 10^{-5}$. Such precision is sufficient since it has to face an experimental uncertainty of the order of $10^{-3}$. We have also verified that the models considered remain perturbative during the running of all couplings. 

In the multi-scale approach, there exist in principle five masses parameters, i.e. $\tilde{m}_{\rm g}$, $\tilde{m}_{\rm W}$, $\tilde{m}_{\rm sq}$, $\tilde{m}_{\rm sl}$ and $\tilde{m}_{\rm h}$, which define the SUSY thresholds. In order to estimate the values for the two masses, $\tilde{m}_{\rm W}$ and $\tilde{m}_{\rm sl}$, assuming $\tilde{m}_{\rm g}/\tilde{m}_{\rm W}=1$ and $\tilde{m}_{\rm sq}/\tilde{m}_{\rm sl}=\tilde{m}_{\bf 10}/\tilde{m}_{\bf 5}=1$ at $M_{\rm GUT}$, we resolve backwards the RGEs of gauginos and sparticles soft masses in the single-scale approach, where $M_{\rm SUSY}$ is of the same order of magnitude of $\tilde{m}_{\rm g}$ and $\tilde{m}_{\rm sq}$, respectively. In particular, concerning matter SUSY particles we assume $\tilde{m}_{\rm sq}/\tilde{m}_{\rm sl}=\tilde{m}_{\rm Q}/\tilde{m}_{\rm L}$ at $M_{\rm SUSY}=\tilde{m}_{\rm sq}$.

Concerning the GUT threshold, for a given value of $\chi_\Sigma$, we iteratively find the value of $\tilde{M}_\Sigma$ through which the chosen value of $\chi_\Sigma$ is achieved.

\section{Results}

In our analysis we try to determine, among the compatible SUSY-GUT models, namely the possible choices $\{\tilde{m}_{\rm h},\,\tilde{m}_{\rm g},\,\tilde{m}_{\rm sq},\,\chi_\Sigma,\,\tan\beta\}$, the energy scale above which SUSY signatures have to show up. To define such a scale, let us consider a given compatible model where the corresponding SUSY particles have  masses $\tilde{m}_{\rm h}$, $\tilde{m}_{\rm g}$, and $\tilde{m}_{\rm sq}$ that admit a minimum. By scanning on all compatible models for fixed $\chi_\Sigma$  we can determine the maximum of previous minima (once that we have marginalized with respect to $\tan\beta$) hereafter denoted by $M_{\rm UB}(\chi_\Sigma)$.
\begin{table}[tbp]
\centering
\begin{tabular}{|l|r|r|}
\hline
Parameters & $\alpha_3(M_{\rm Z})$ & $M_{\rm GUT}$  \\
\hline
\hline
$\chi_\Sigma$: $1\rightarrow10$ & +2.0 \% &  +106 \%  \\
\hline
$\tilde{m}_{\rm g}$: $1\rightarrow10$ TeV & -2.2 \% & -47 \%  \\
\hline
$\tilde{m}_{\rm sq}$: $1\rightarrow10$ TeV & +0.2 \% & -2.6 \%  \\
\hline
$\tilde{m}_{\rm h}$: $1\rightarrow10$ TeV & -4.5 \% & -16 \%  \\
\hline
\end{tabular}
\caption{\label{tab:behaviors}Level of dependence of $\alpha_3(M_{\rm Z})$ and $M_{\rm GUT}$ on the main parameters related to SUSY and GUT thresholds. The dependence on $\tan\beta$, at this level, results to be negligible.}
\end{table}

In Table \ref{tab:behaviors} we report the behaviors of $\alpha_3(M_{\rm Z})$ and $M_{\rm GUT}$ as a function of the SUSY and GUT thresholds. In particular, concerning $\alpha_3(M_{\rm Z})$ one observes two opposite dependences on the thresholds. The increasing in the GUT threshold, $\chi_\Sigma$, produces an analogous increasing in  $\alpha_3(M_{\rm Z})$, whereas the opposite occurs for SUSY thresholds. This can lead to a possible balancing between these different effects that allows for larger masses in the SUSY spectrum. As it appears from the Table, the squarks provides a negligible contribution to $\alpha_3(M_{\rm Z})$ hence justifying the simplicity ansatz we already quoted $\tilde{m}_{\bf{ 10}}/\tilde{m}_{\bf 5} =1$ at $M_{\rm GUT}$. The behavior of $M_{\rm GUT}$ on the thresholds is qualitatively similar, but the strong dependence on $\chi_\Sigma$ makes more difficult a possible balancing.

\begin{figure}[h!]
\centering
\begin{tabular}{|c|c|c|}
\hline
$\chi_\Sigma=1$ $\left( \lambda_\Sigma \sim 0.5 \right)$  & $\chi_\Sigma=3$ $\left( \lambda_\Sigma \sim 0.2 \right)$  & $\chi_\Sigma=10$ $\left( \lambda_\Sigma \sim 0.05 \right)$ \\
\hline
&&\\[-10pt]
\includegraphics[scale=0.178]{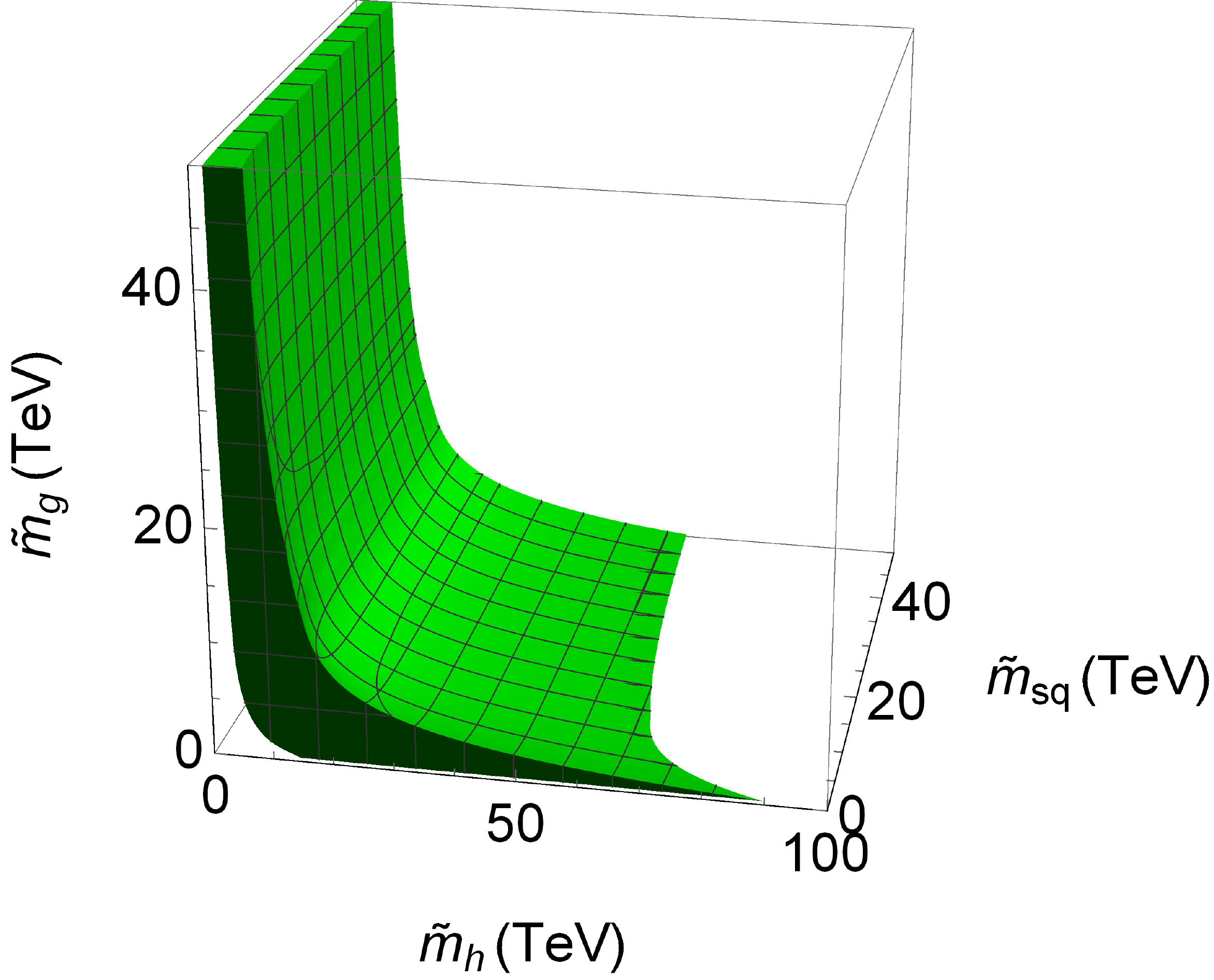} &
\includegraphics[scale=0.178]{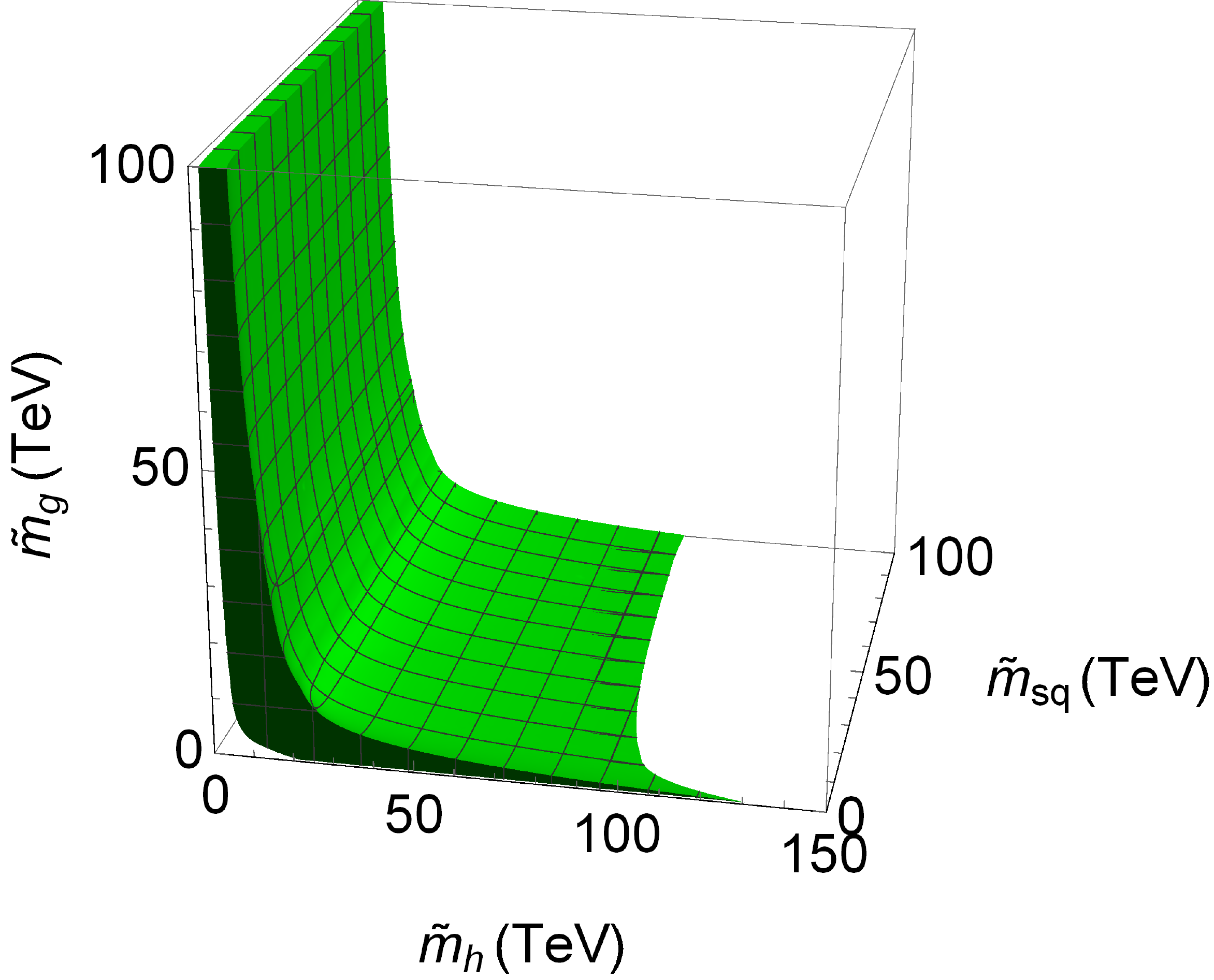} &
\includegraphics[scale=0.178]{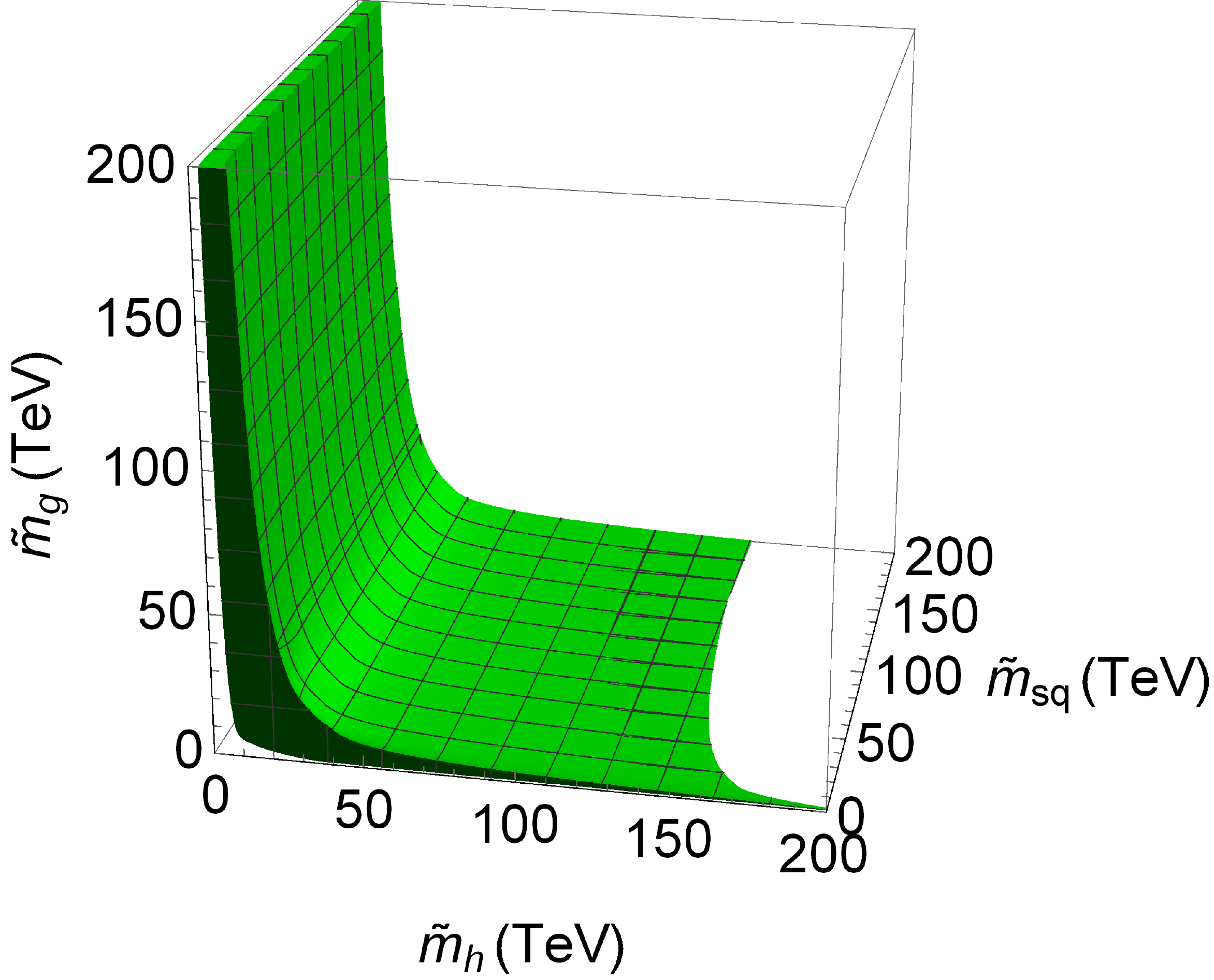} \\
\hline
&&\\[-10pt]
\includegraphics[scale=0.178]{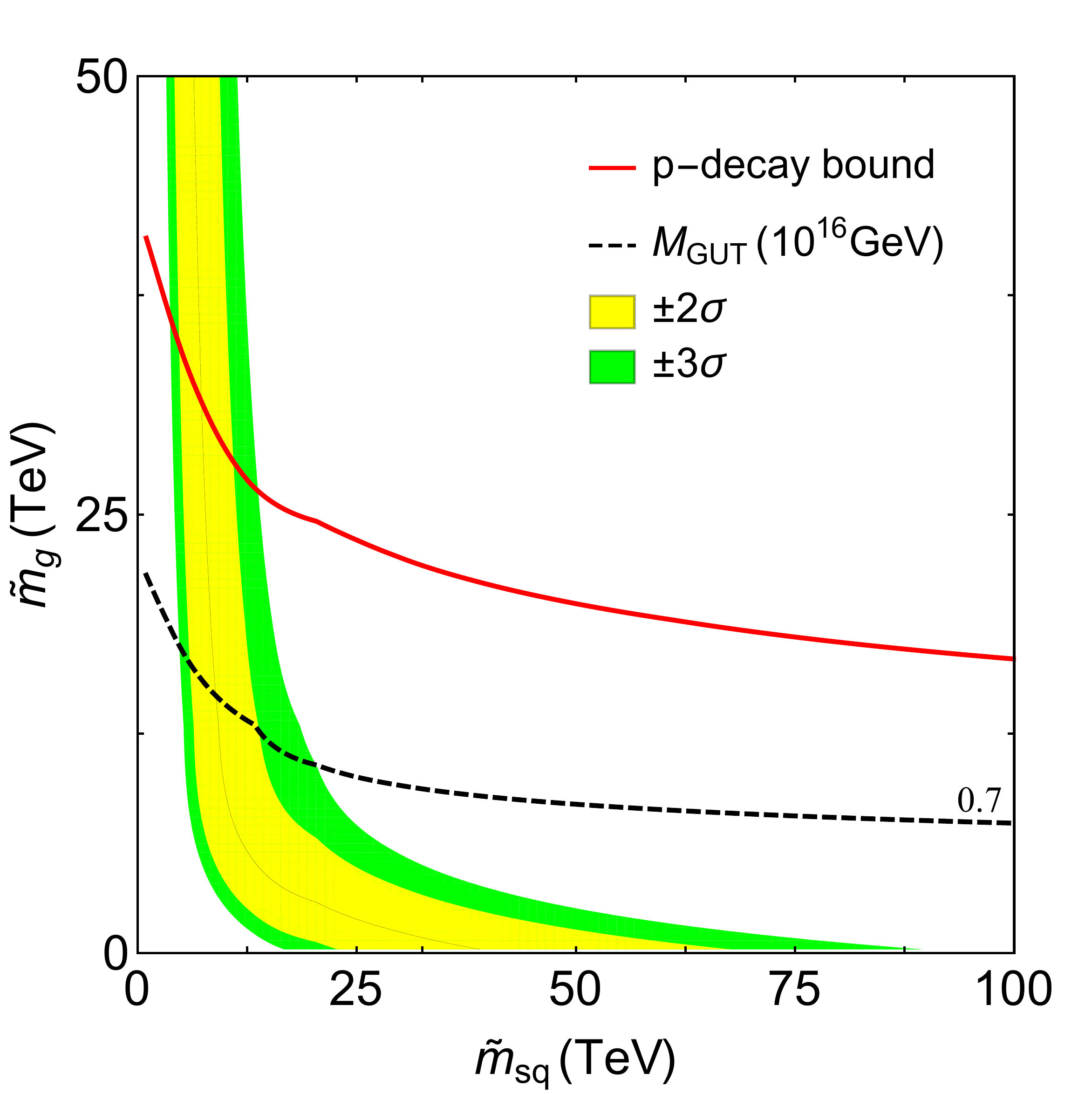} &
\includegraphics[scale=0.178]{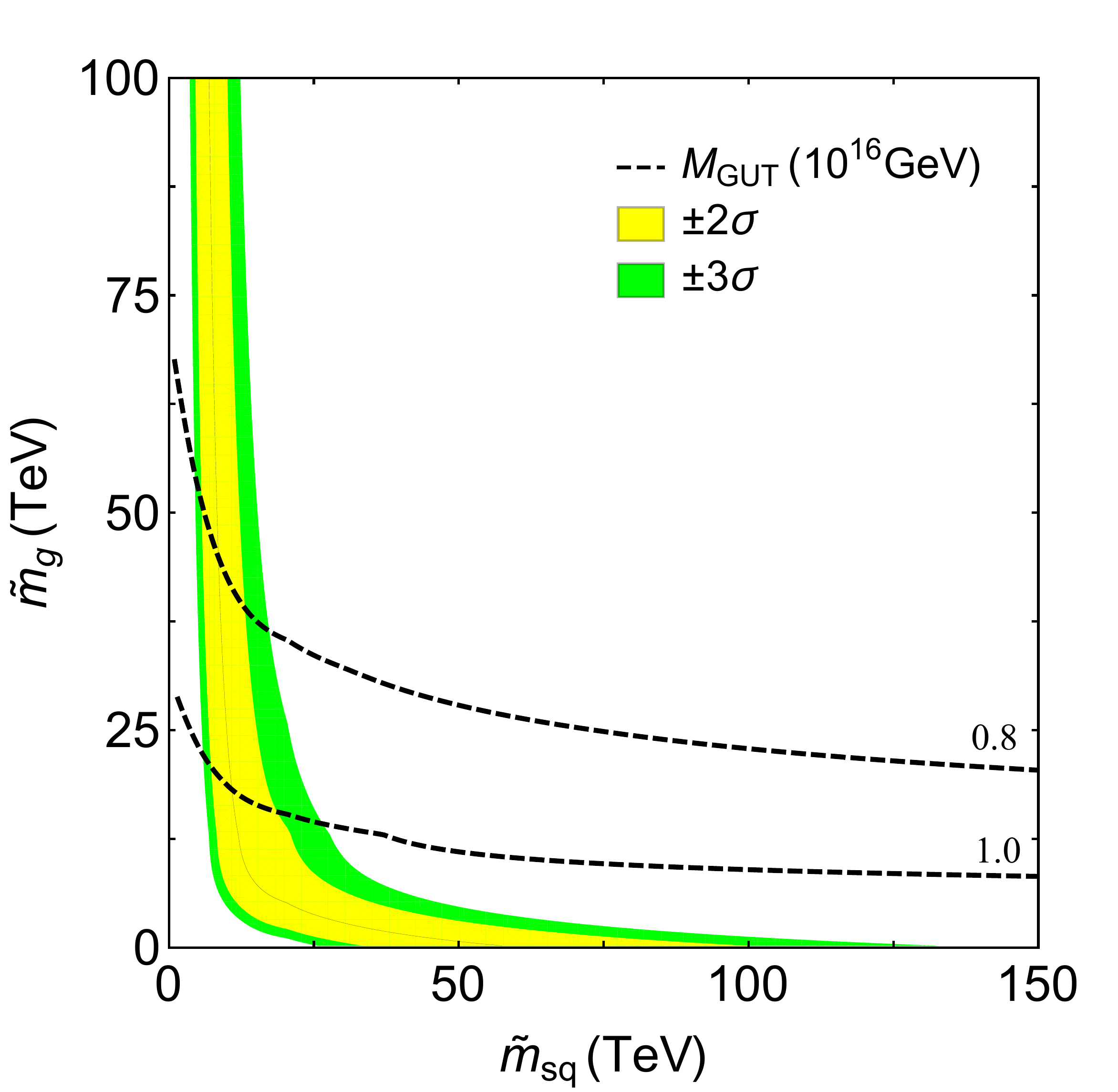} &
\includegraphics[scale=0.178]{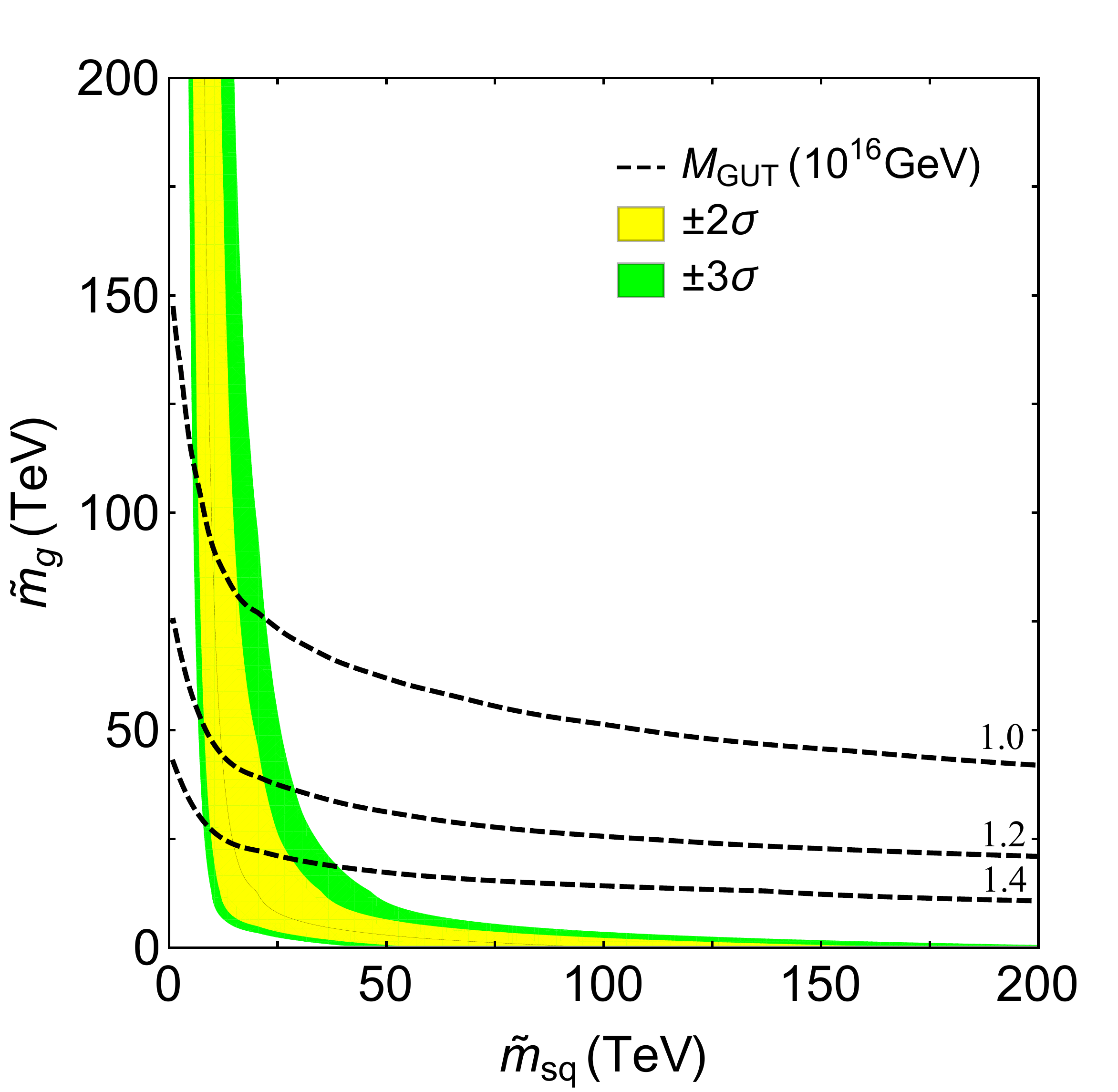} \\
{\small (a) $\tilde{m}_{\rm h}=\tilde{m}_{\rm sl}$} & {\small (b) $\tilde{m}_{\rm h}=\tilde{m}_{\rm sl}$} & {\small (c) $\tilde{m}_{\rm h}=\tilde{m}_{\rm sl}$} \\
\hline
&&\\[-10pt]
\includegraphics[scale=0.178]{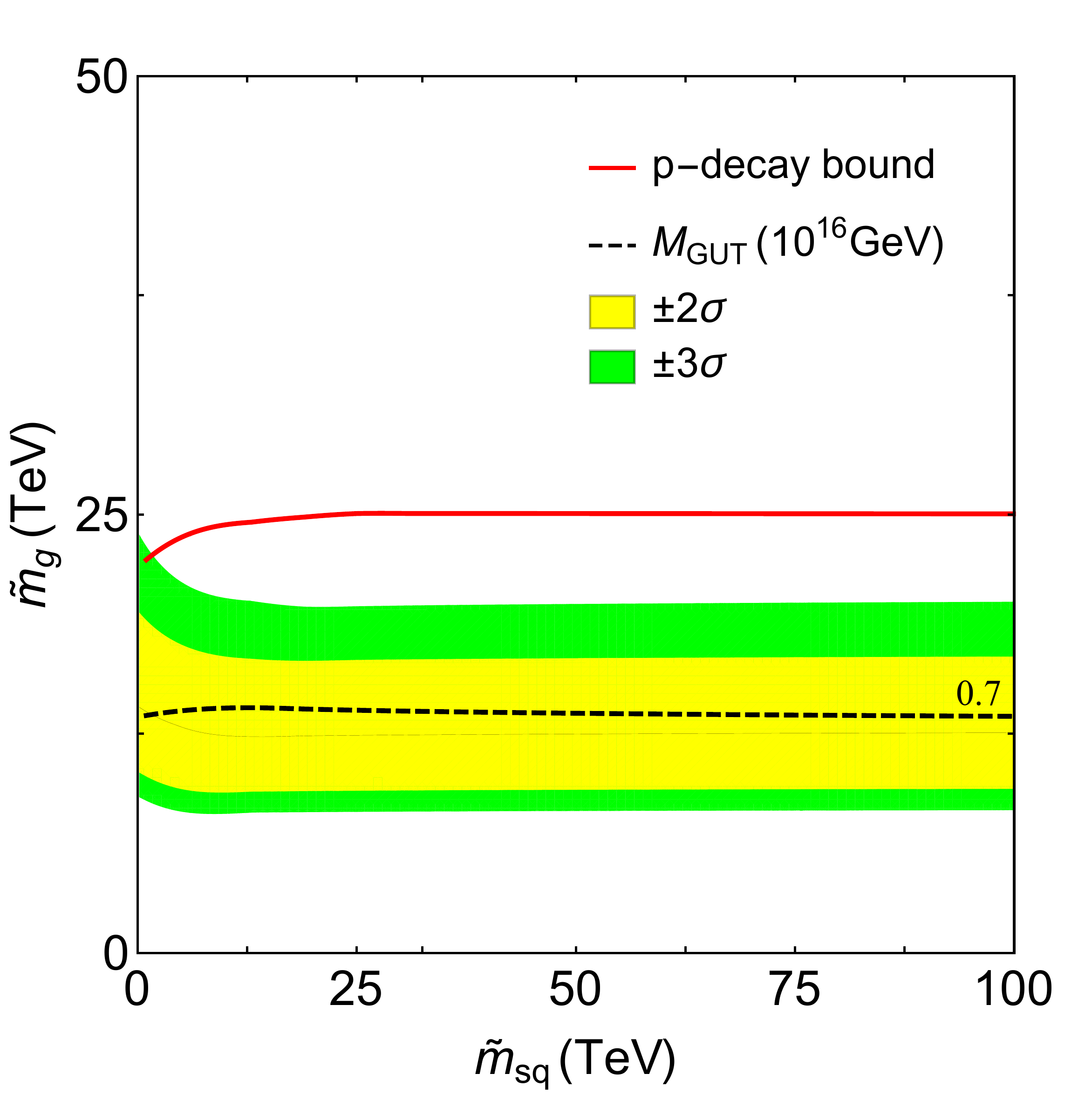} &
\includegraphics[scale=0.178]{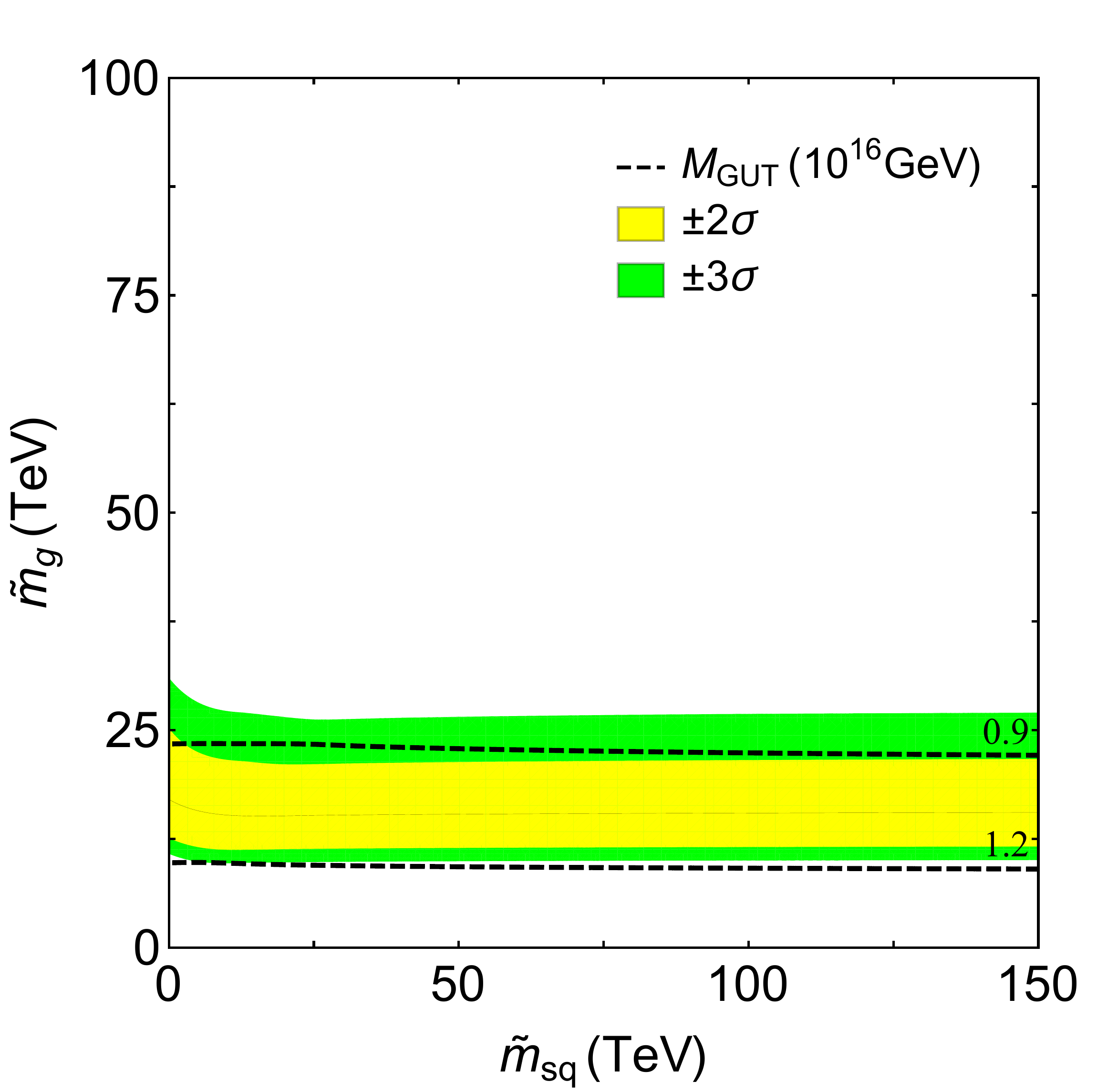} &
\includegraphics[scale=0.178]{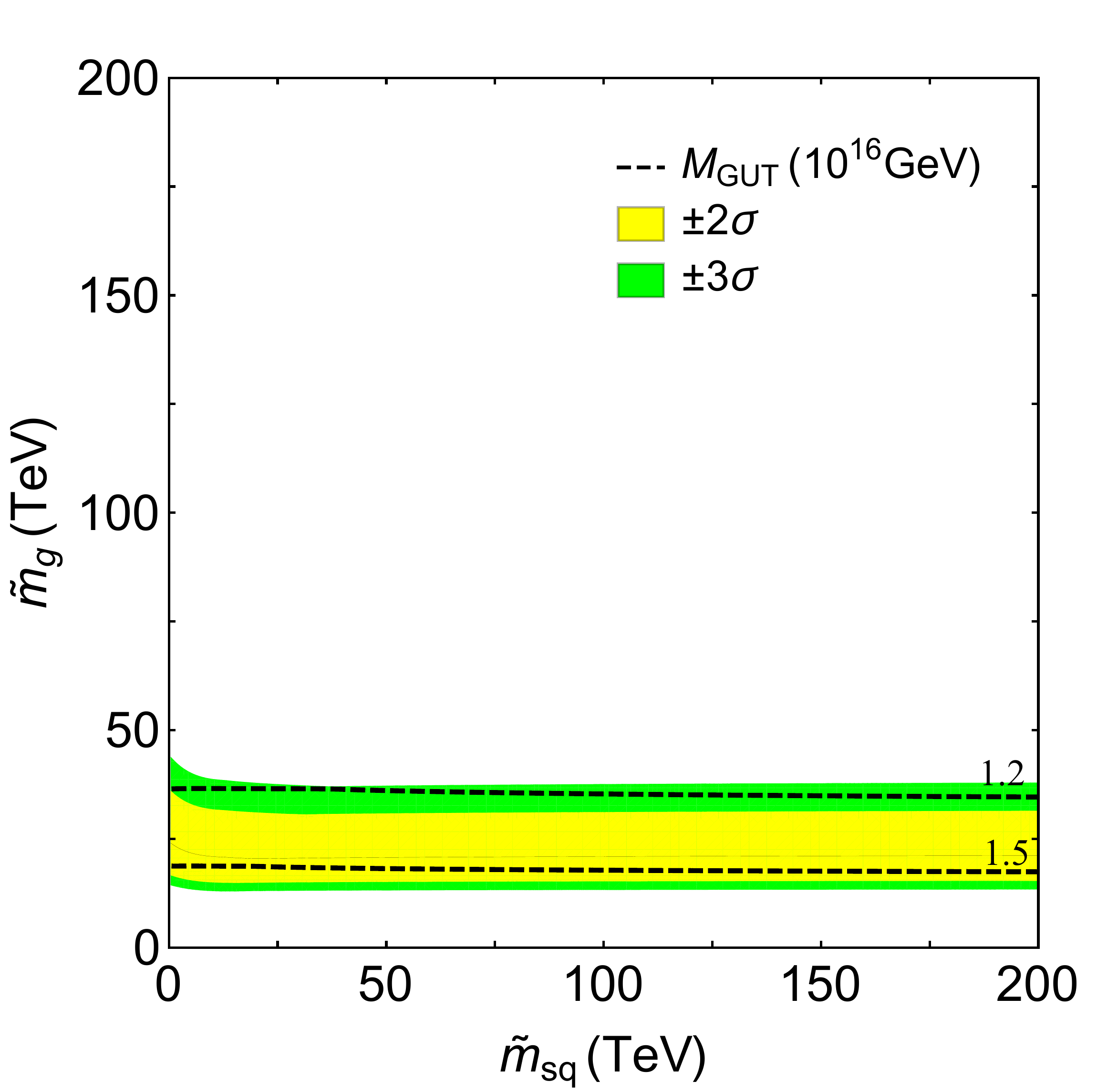} \\
{\small (d) $\tilde{m}_{\rm h}=\tilde{m}_{\rm W}$} & {\small (e) $\tilde{m}_{\rm h}=\tilde{m}_{\rm W}$} & {\small (f) $\tilde{m}_{\rm h}=\tilde{m}_{\rm W}$} \\
\hline
&&\\[-10pt]
\includegraphics[scale=0.178]{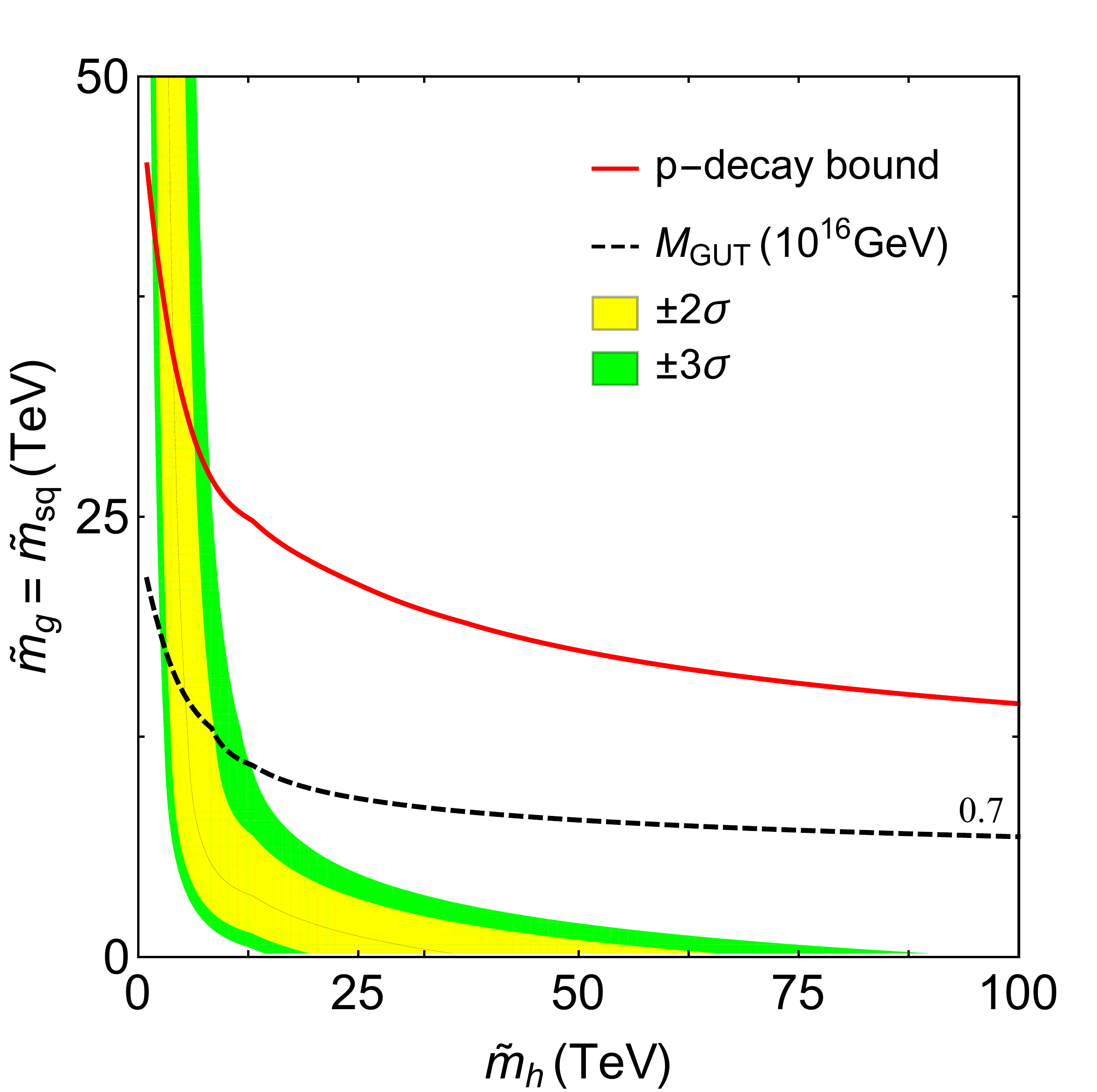} &
\includegraphics[scale=0.178]{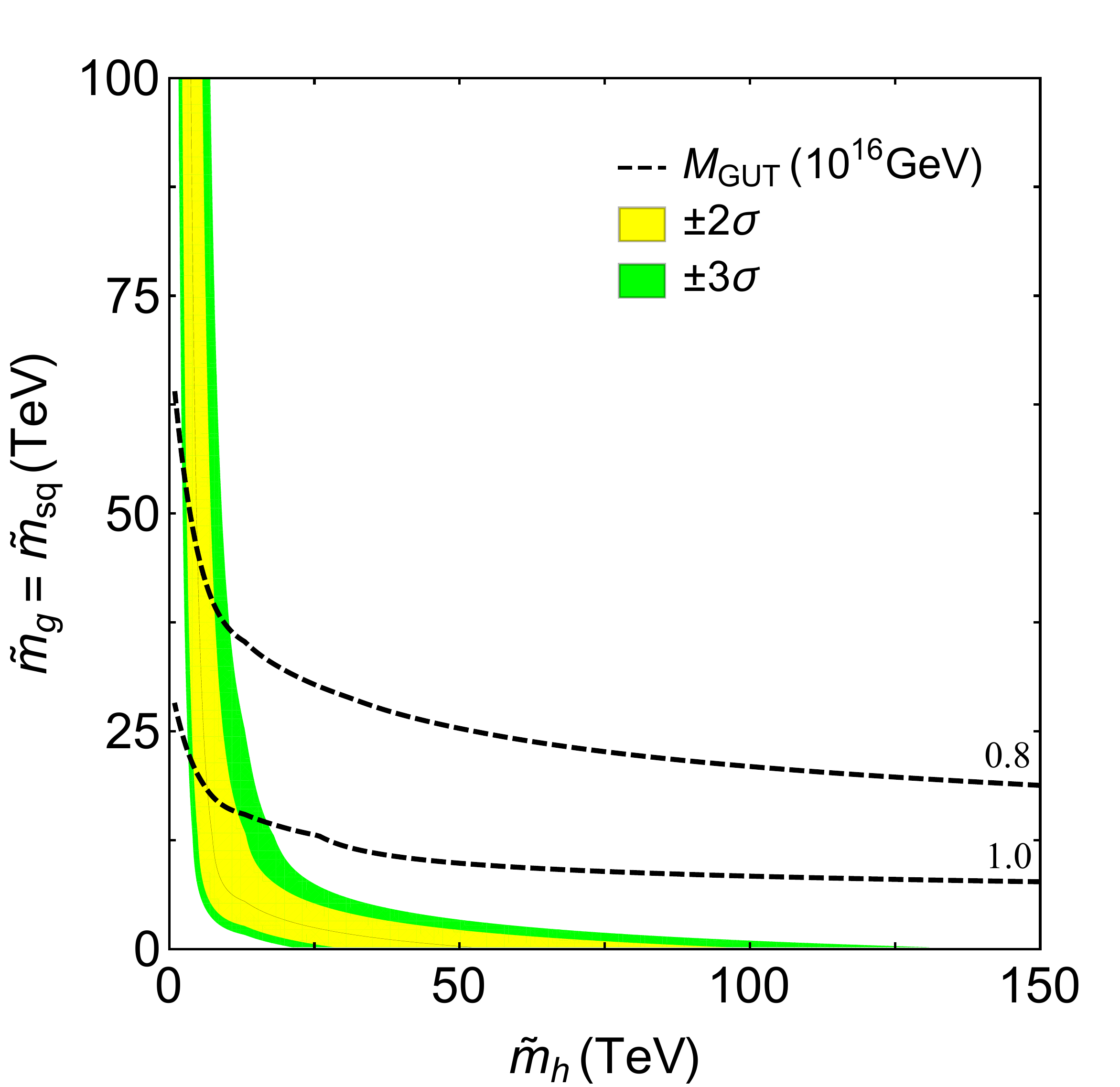} &
\includegraphics[scale=0.178]{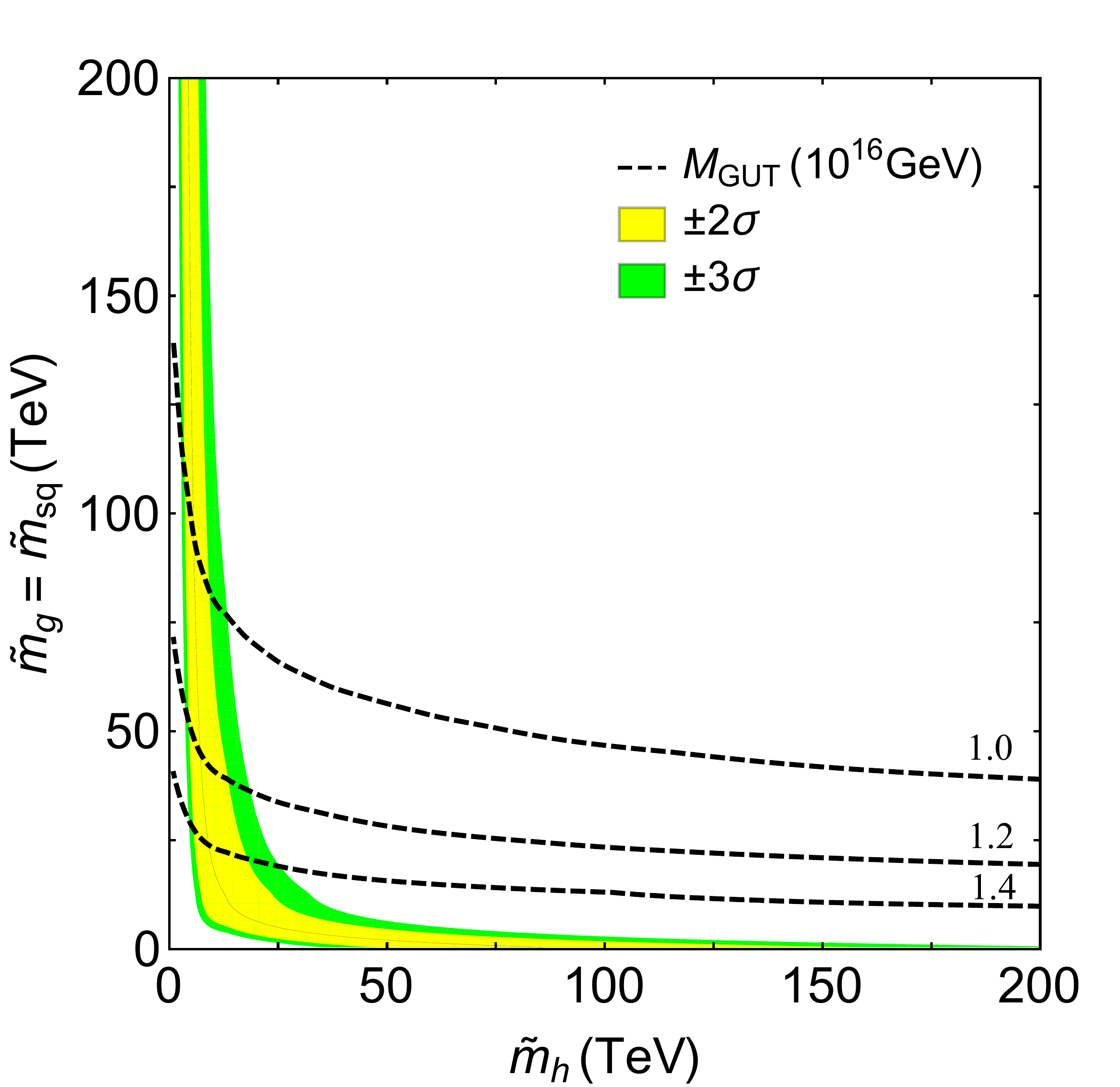} \\
{\small (g) $\tilde{m}_{\rm g}=\tilde{m}_{\rm sq}$} & {\small (h) $\tilde{m}_{\rm g}=\tilde{m}_{\rm sq}$} & {\small (i) $\tilde{m}_{\rm g}=\tilde{m}_{\rm sq}$} \\
\hline
\end{tabular}
\caption{In the first row of the panel the compatibility regions for the masses $\tilde{m}_{\rm h}$, $\tilde{m}_{\rm g}$, and $\tilde{m}_{\rm sq}$ for different values of  $\chi_\Sigma$ are reported. The second, third and fourth row provide the allowed regions  once they have been assumed the relations $\tilde{m}_{\rm h}=\tilde{m}_{\rm sl}$, $\tilde{m}_{\rm h}=\tilde{m}_{\rm W}$ and $\tilde{m}_{\rm g}=\tilde{m}_{\rm sq}$, respectively (green and yellow regions correspond to $99\%$ and $95\%$ CL respectively). The dashed black lines correspond to given values of $M_{\rm GUT}$ expressed in $10^{16}$ GeV, whereas the red line, if present, bounds from above the region allowed by proton lifetime (Eq. \eqref{eq:proton}).
\label{fig:plots}}
\end{figure}

In Fig. \ref{fig:plots} we report on the first row of the panel the compatibility region for the masses $\tilde{m}_{\rm h}$, $\tilde{m}_{\rm g}$,  $\tilde{m}_{\rm sq}$ for different values of GUT threshold $\chi_\Sigma$. As it is clear, $\tilde{m}_{\rm h}$ and $\tilde{m}_{\rm g}$ result to have a reasonable inverse correlation and this occurs almost independently of $\tilde{m}_{\rm sq}$. From the Figure one can observe the presence of an upper bound for $\tilde{m}_{\rm h}$ that is smaller than the possible ones for the other particles, and that such value appears to be a monotonic increasing function of $\chi_\Sigma$. Note that for simplicity in these pictures we do not superimpose the constraints coming from proton lifetime and $b$-$\tau$ unification.

\begin{figure}[t!]
\centering
\begin{tabular}{|c|c|c|}
\hline
$\chi_\Sigma=1$ $\left( \lambda_\Sigma \sim 0.5 \right)$ & $\chi_\Sigma=3$ $\left( \lambda_\Sigma \sim 0.2 \right)$  & $\chi_\Sigma=10$ $\left( \lambda_\Sigma \sim 0.05 \right)$  \\
\hline
&&\\[-10pt]
\includegraphics[width=0.3\textwidth]{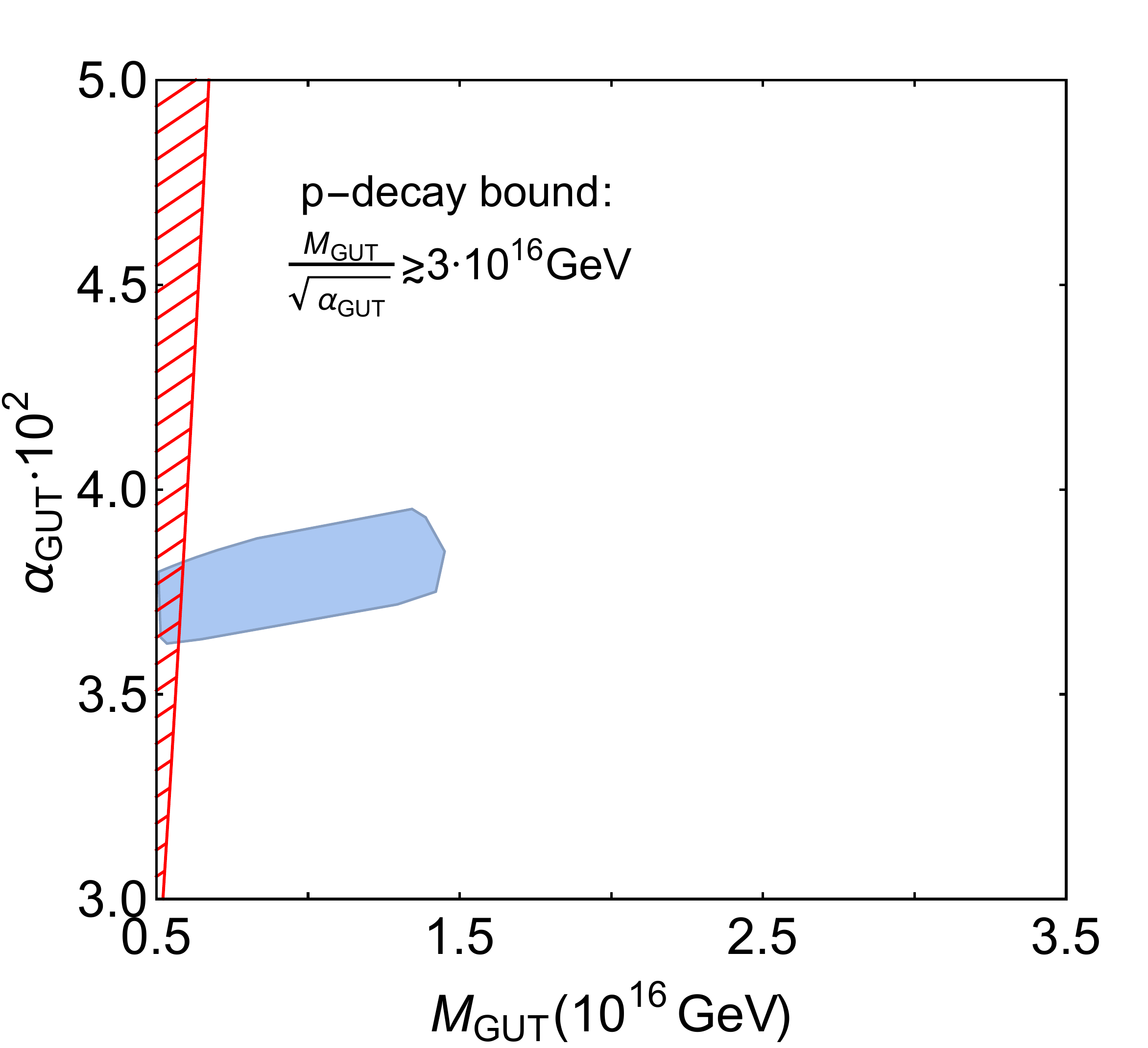} &
\includegraphics[width=0.3\textwidth]{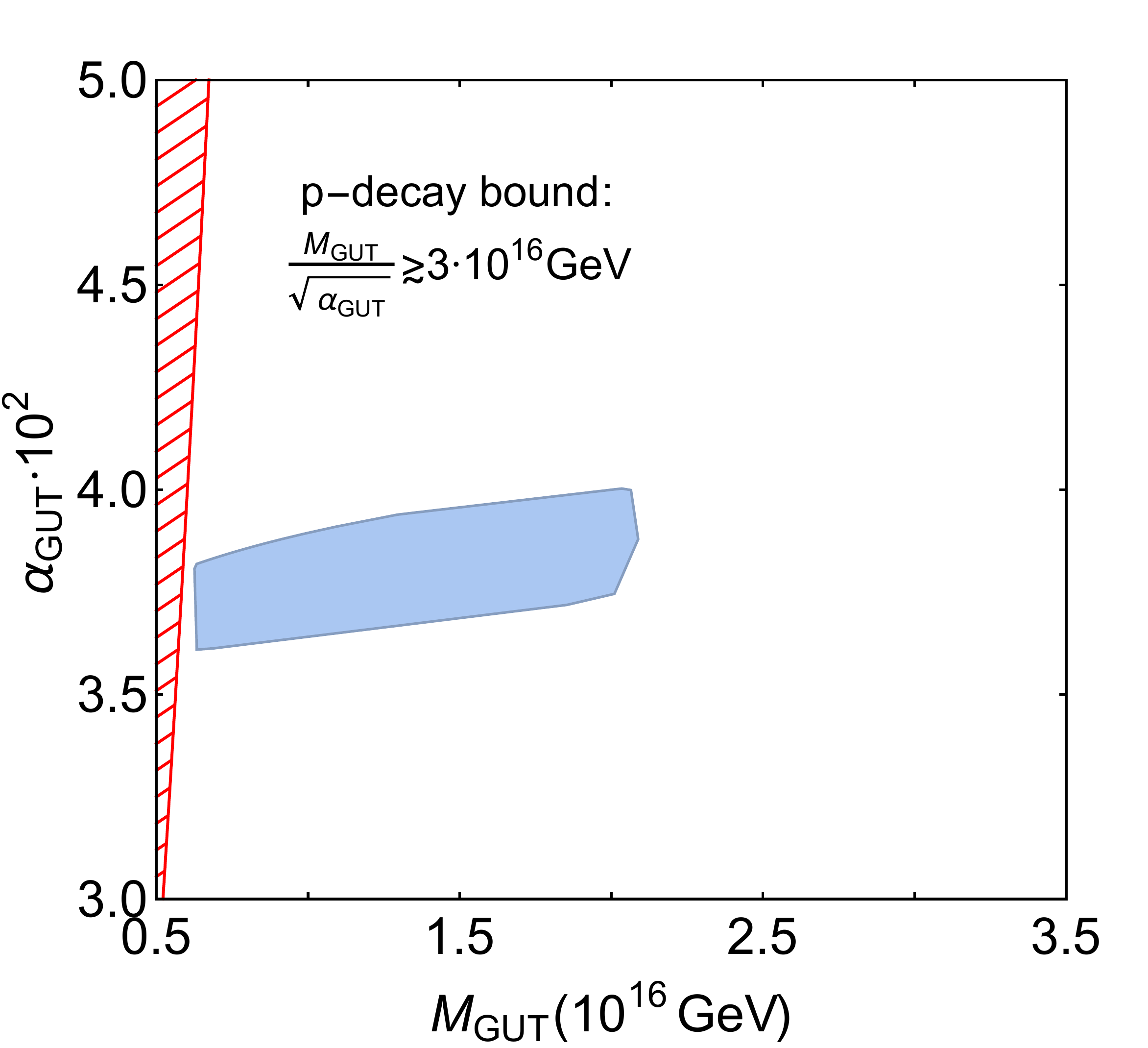} &
\includegraphics[width=0.3\textwidth]{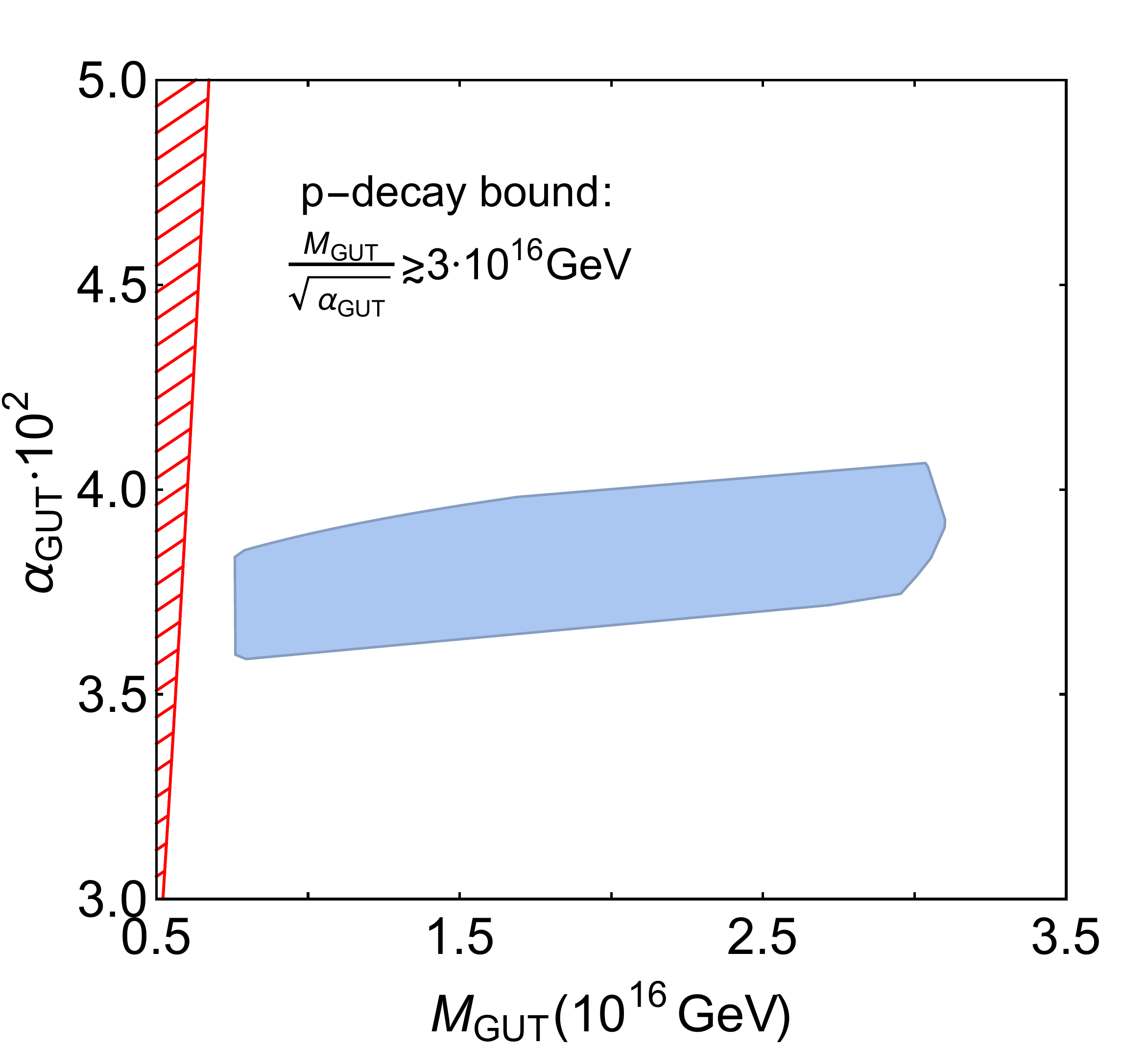} \\
\hline
\end{tabular}
\caption{\label{fig:GUT_region}Allowed regions in the $M_{\rm GUT}$-$\alpha_{\rm GUT}$ plane obtained for different values of GUT threshold $\chi_\Sigma$. The red-dashed regions are excluded by proton lifetime constraint of Eq.~\eqref{eq:proton}.}
\end{figure}

The other rows of Fig. \ref{fig:plots} provide the allowed regions of the input parameters once one has reduced the level of arbitrariness among the values of SUSY masses. In particular in the second, third and fourth row one assumes the relations $\tilde{m}_{\rm h}=\tilde{m}_{\rm sl}$, $\tilde{m}_{\rm h}=\tilde{m}_{\rm W}$ and $\tilde{m}_{\rm g}=\tilde{m}_{\rm sq}$, respectively. Such situations cannot be straightforwardly derived by the 3d plots of the first row, and hence they complement such information. Moreover, these three different situations are in agreement with the fine tuning of Eq. \eqref{eq:Higgs_finetunig}, which require the masses $\tilde{m}_{\rm g}$ (SSB F-terms), $\tilde{m}_{\rm sq}$ (SSB D-terms) and $\tilde{m}_{\rm h}$ (supersymmetric $\mu$-term) to be of the same order of magnitude. The superimposed dashed lines correspond to given values of  $M_{\rm GUT}$, expressed in $10^{16}$ GeV, whereas  the red line, if present, bounds from above the region allowed by proton lifetime. Note that Eq. \eqref{eq:proton} bounds the compatibility regions for small values of $\chi_\Sigma$ only. This is due to the fact that larger values of  $\chi_\Sigma$ are able to spread and shift up to larger values the compatibility regions for SUSY particle masses (see second and third column of Fig. \ref{fig:plots}). Moreover $\chi_\Sigma$ affects the $M_{\rm GUT}$ allowed region as well, as it is clearly shown in Fig. \ref{fig:GUT_region}, where it can be seen how increasing the values of $\chi_\Sigma$ one gets larger values for the unification scale. The red-dashed regions of Fig. \ref{fig:GUT_region} are excluded by proton lifetime constraint of Eq. \eqref{eq:proton}. 

The anti-correlation showed by some of the mass parameters suggests that to look for $M_{\rm UB}(\chi_\Sigma)$, it is convenient to assume particular relations among SUSY particle masses. In particular, in the $\tilde{m}_{\rm h}$-$\tilde{m}_{\rm g}$ plane one can see that if a mass is very light, the other one has to be very heavy, whereas the quantity $\tilde{m}_{\rm sq}$ is not bound at all. This suggests that $M_{\rm UB}(\chi_\Sigma)$ can be found by imposing $\tilde{m}_{\rm h}=\tilde{m}_{\rm g}=\tilde{m}_{\rm sq}$ in the allowed region, and looking for the maximum of such values.
\begin{figure}[t!]
\centering
\includegraphics[scale=0.27]{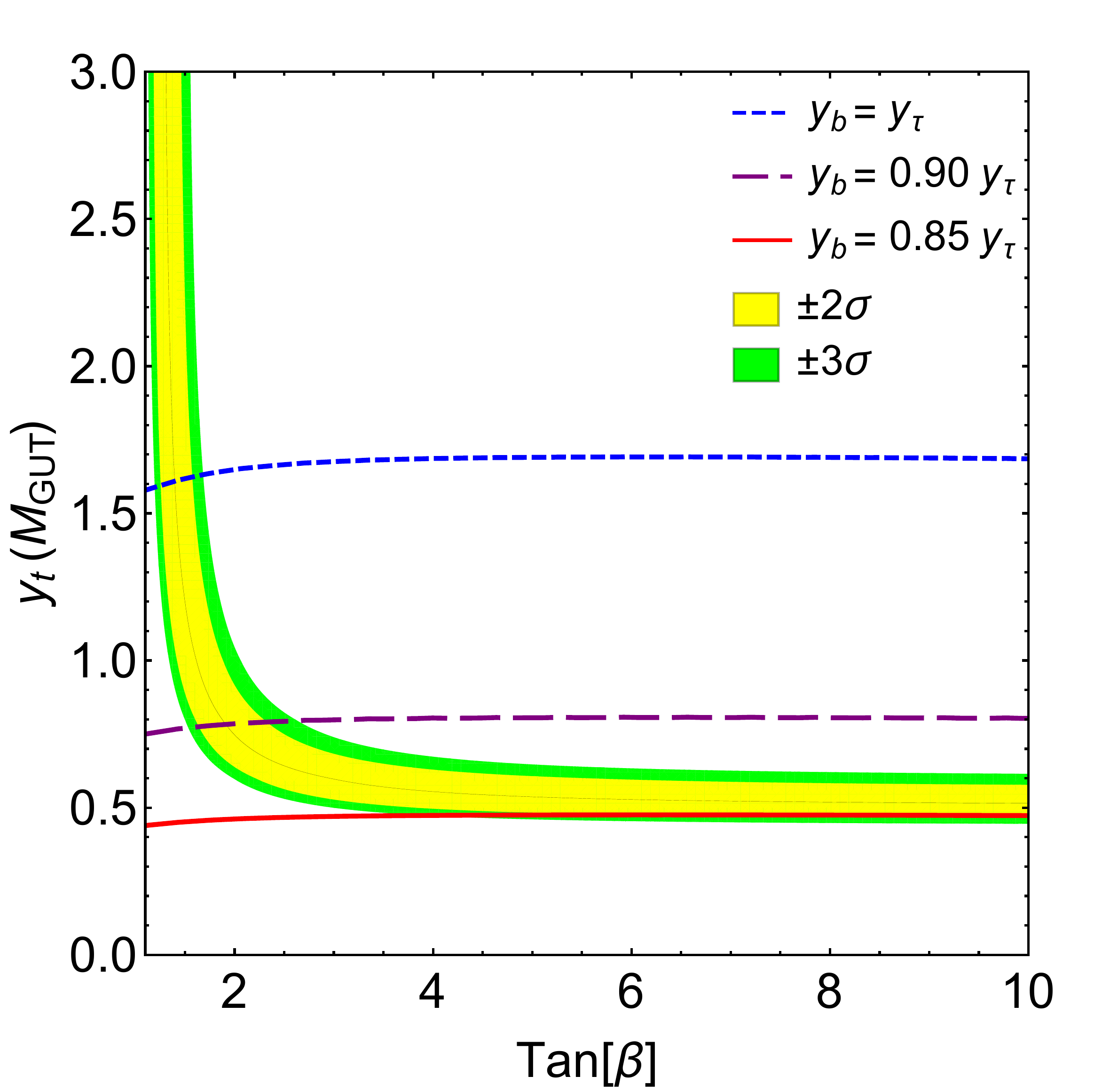}
\hspace{5.mm}
\includegraphics[scale=0.285]{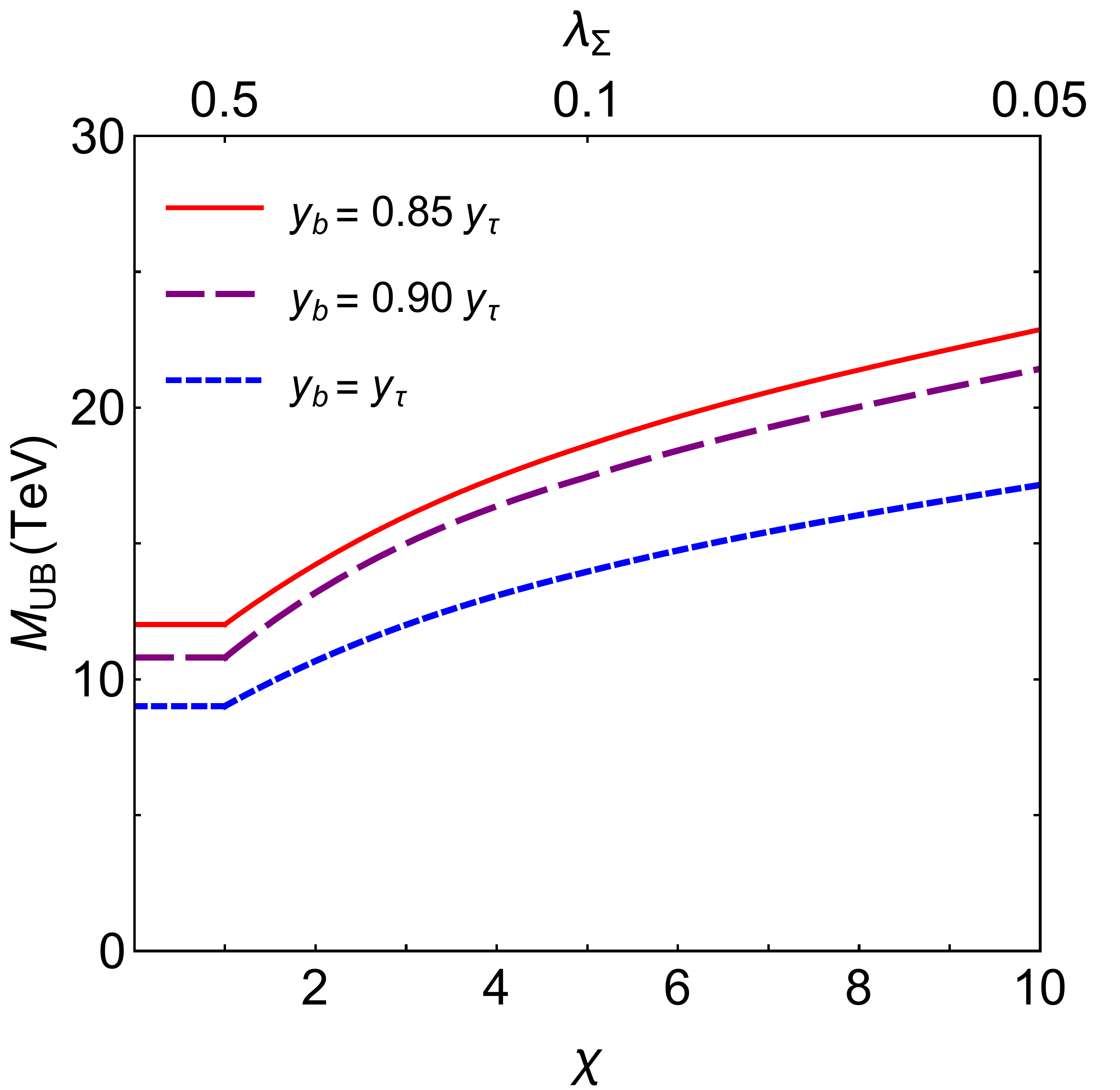}
\\(a) \hskip67.mm (b)\\
\caption{\label{fig:worst} (a) Allowed region in the plane $\tan\beta$-$y_{\rm t}(M_{\rm GUT})$. The yellow (green) region corresponds to compatibility within $2\sigma$ ($3\sigma$) uncertainty. The blue, purple and red lines bound from below the allowed region once that exact or partial $b$-$\tau$ unification at $M_{\rm GUT}$ is respectively assumed (see Eq.~\eqref{eq:btau}). (b) The quantity $M_{\rm UB}$ as a function of $\chi_\Sigma$. On the upper part of the picture, we report the values of $\lambda_\Sigma$ corresponding to the values of $\chi_\Sigma$ according to Eq. \eqref{eq:chi_GUT}.}
\end{figure}

In the left panel of Fig. \ref{fig:worst} we provide the allowed region in the plane $\tan\beta$-$y_{\rm t}(M_{\rm GUT})$. As it is clear from the plot, for moderate $\tan\beta$ ({\it naturalness} requirement, namely small fine tuning in the Higgs and Yukawa sector) $y_{\rm t}(M_{\rm GUT})$ is bound to be larger than $0.5$. In the case of exact $(y_b=y_\tau)$ or partial ($y_b=0.90 \, y_\tau$ and $y_b=0.85 \, y_\tau$) $b$-$\tau$ unification at $M_{\rm GUT}$, according to Eq.~\eqref{eq:btau}, the quantity $y_t(M_{\rm GUT})$ cannot be lower than $\sim 1.5$, $\sim 0.7$ and $\sim 0.5$, respectively. It is worth observing, that the $b$-$\tau$ unification at GUT energy also imposes $\tan\beta\lesssim 3.0$ for the case $y_b=0.90 \, y_\tau$. However, larger values for $\tan\beta$ can be obtained by relaxing the relation of Eq.~\eqref{eq:btau}. In particular, the allowed region in the plane $\tan\beta$-$y_{\rm t}(M_{\rm GUT})$ is not bounded at all for $y_b=0.80 \, y_\tau$, which is considered as the case ``without $b$-$\tau$ unification".

In the right panel of Fig. \ref{fig:worst} we show the quantity $M_{\rm UB}(\chi_\Sigma)$ as function of GUT threshold $\chi_\Sigma$. On the upper part of the picture we report the values of $\lambda_\Sigma$ corresponding to the particular values of $\chi_\Sigma$. As it can be easily seen, too large values of $\chi_\Sigma$ would correspond to excessively small values of $\lambda_\Sigma$ that can be considered {\it unnatural}. Moreover, to assume large values for $\chi_\Sigma > 10$ would also imply $M_{\rm GUT}$ unnaturally approaching Planck scale. For these reasons in our analysis we bound the values of $\chi_\Sigma$ in the conservative interval $1 \div 10$. As shown in the plot, $b$-$\tau$ unification at $M_{\rm GUT}$ prefers lighter SUSY particles, implying lower values for $M_{\rm UB}(\chi_\Sigma)$.

For all these reasons, one gets that the {\it upper bound} for $M_{\rm UB}(\chi_\Sigma)$ is  $\sim 20$ TeV whenever $\chi_\Sigma \leq 10$. For energy scale larger than such {\it upper bound} a generic SUSY-GUT model satisfying our requirements has to show up  via the detection of Higgsinos or gluinos. Such a prediction cannot be extended to squarks that could be much heavier. 

Note that, in case we had used more stringent bounds due to dimension-5 operators mediating proton decay (even though highly model dependent), we would have excluded larger portions of the allowed regions reported in Fig. \ref{fig:GUT_region} thus favoring larger values of $\chi_\Sigma$. In this case, the determination of the upper bound of $\sim 20$ TeV for $M_{\rm UB}$ would remain completely unchanged since it only depends on the largest values of $\chi_\Sigma$ compatible with our requirements.

\section{Conclusions}

Besides the recent detection of a new particle compatible with the Higgs boson that completes the spectacular set of experimental evidences supporting the SM, it is hardly to believe that a QFT based on the gauge group $SU(3)\times SU(2)  \times U(1)$ can represent the deepest  description of fundamental interactions. This consideration mainly follows from the observation that  several phenomena or open problems suggest the presence of physics beyond the SM. Among them, the hint of unification of all gauge couplings for extreme large energy, and the problem related to a {\it natural} separation of very different energy scales in a field theory with scalars ({\it hierarchy problem}) strongly indicate the need for more profound schemes.

 The GUT paradigm once implemented in a SUSY framework is able to simply address all these problems. Nevertheless, since the 8 TeV LHC run I has greatly constrained SUSY-GUT, one can ask if such simple scenario is still viable. In the present analysis, by considering a generic SUSY-GUT obeying quite general assumptions ({\it SU(5) bottleneck}; consistency with third family fermion masses and with the experimental limit on proton decay; absence of special fine tunings among couplings, which are ${\cal O}(1)$ at $M_{\rm GUT}$) we have analyzed the limits on the mass spectrum of SUSY particles once the experimental constraints at $M_Z$ and proton decay limit are considered.

Parametrizing the SUSY mass spectrum in terms of three quantities related to SUSY soft masses $\tilde{m}_{\rm g}$ (SSB F-terms) for gluinos, $\tilde{m}_{\rm sq}$ (SSB D-terms) for squarks, and $\tilde{m}_{\rm h}$ (supersymmetric $\mu$-term) for Higgsinos, we have looked at those values compatible with the previous requirements. The aim was the determination of the {\it upper bound} for the minima of compatible spectra. Such energy scale defines a threshold above which SUSY signatures have to show up. This study provides indications about the chances to unveil new physics on LHC run II and on future colliders.

In order to be more conservative in our prediction we have discussed the general possibility that SUSY-GUT posses several thresholds. This means to assume possible splittings between the masses of SUSY particles (multi-scale approach), and analogously, to admit that some GUT particles can take mass below the unification scale (GUT thresholds). In comparison with the single-scale approach where the SUSY breaking scale is of order of few TeV, in the multi-scale scenario we find that SUSY particles masses can be much heavier.

We claim that if a SUSY-GUT model is the proper way to describe physics beyond the SM and under the {\it natural} assumptions reported above, the lightest gluino or Higgsino cannot have a mass larger than $\sim$ 20 TeV. The requirement of $b$-$\tau$ unification at GUT energy scale slightly reduces such {\it upper bound} of few TeV. It is worth observing that our conclusions are strongly affected by the experimental uncertainty on $\alpha_3(M_{\rm Z})$ and on the proton lifetime lower limit. According to these results a new generation of colliders able to achieve the $100$ TeV \cite{Gomez-Ceballos:2013zzn, fcc_hh} energy range would have the chance to cover almost all the parameter space beneath a {\it natural} unifying scheme.

\section*{\bf Acknowledgments} 

We acknowledge support of the MIUR grant for the Research Projects of National Interest  PRIN 2012 No. 2012CPPYP7 ``Astroparticle Physics". The work of Z.B. was supported in part by Rustaveli National Science Foundation grant No. DI/8/6-100/12.

\end{document}